\documentclass[11pt,a4paper]{article}
\pdfoutput=1
\usepackage{jheppub}
\usepackage{amsthm,amsbsy,amsfonts,mathrsfs,enumerate,float,wrapfig,amsmath}

\newcommand{\be}{\begin{equation}}
\newcommand{\ee}{\end{equation}}
\newcommand{\bea}{\begin{eqnarray}}
\newcommand{\eea}{\end{eqnarray}}

\title{Equivalence of several descriptions for 6d SCFT}

\author[a, b]{Hirotaka Hayashi,}
\author[c, d]{Sung-Soo Kim,}
\author[d]{Kimyeong Lee,}
\author[d]{and Futoshi Yagi}

\affiliation[a]{Tokai University, 4-1-1 Kitakaname, Hiratsuka-shi, Kanagawa 259-1292, Japan}
\affiliation[b]{Departamento de F\'isica Te\'orica and Instituto de F\'isica Te\'orica UAM/CSIC,\\ Universidad Aut\'onoma de Madrid, Cantoblanco, 28049 Madrid, Spain}
\affiliation[c]{School of Physical Electronics, University of Electronic Science and Technology of China, North Jianshe Road,Chengdu, 611731, China}
\affiliation[d]{Korea Institute for Advanced Study, 
85 Hoegi-ro Dongdaemun-gu, Seoul, 02455, Korea}

\emailAdd{h.hayashi@tokai.ac.jp}
\emailAdd{sungsoo.kim@uestc.edu.cn}
\emailAdd{klee@kias.re.kr}
\emailAdd{fyagi@kias.re.kr}

\abstract{
We show that the three different looking BPS partition functions, namely the elliptic genus of the 6d $\mathcal{N}=(1,0)$ $Sp(1)$ gauge theory with $10$ flavors and a tensor multiplet, the Nekrasov partition function of the 5d $\mathcal{N}=1$ $Sp(2)$ gauge theory with $10$ flavors, and the Nekrasov partition function of the 5d $\mathcal{N}=1$ $SU(3)$ gauge theory with $10$ flavors, are all equal to each other under specific maps among gauge theory parameters. This result strongly suggests that the three gauge theories have an identical UV fixed point. Type IIB 5-brane web diagrams play an essential role to compute the $SU(3)$ Nekrasov partition function as well as establishing the maps.
}

\begin{document}
\preprint{
\begin{flushright}
\tt 
KIAS-P16048\\
\end{flushright}
}

\maketitle


\section{Introduction}
\label{sec:intro}

It has been discussed that some five-dimensional (5d) $\mathcal{N}=1$ supersymmetric gauge theories have their ultraviolet (UV) fixed points where the gauge theories become interacting superconformal field theories \cite{Seiberg:1996bd, Morrison:1996xf, Douglas:1996xp, Intriligator:1997pq}
even though they are perturbatively non-renormalizable.
Such superconformal field theories are often addressed via brane configurations or Calabi-Yau compactifications
in string theory or M-theory.
For example, based on the existence of consistent $(p,q)$ 5-brane configurations
developed in \cite{Aharony:1997ju, Aharony:1997bh, Benini:2009gi}, 
it is conjectured that the 5d $\mathcal{N}=1$ $SU(N)$  
gauge theory with $N_f \le 2N+3$ flavors \cite{Bergman:2014kza, Kim:2015jba, Hayashi:2015fsa, Yonekura:2015ksa}
as well as the 5d $\mathcal{N}=1$ $Sp(N-1)$
gauge theory with $N_f \le 2N+3$ flavors \cite{Bergman:2015dpa}
have UV fixed points, 
some of which exceed the bound discussed in \cite{Intriligator:1997pq}.

Based on Tao diagrams \cite{Kim:2015jba, Hayashi:2015fsa} and the instanton operator analysis \cite{Yonekura:2015ksa}, 
it is further discussed that the 5d $\mathcal{N}=1$ $SU(N)$ supersymmetric gauge theory with $N_f = 2N+4$ flavors have a UV fixed point, where a six-dimensional (6d) superconformal field theory named as ``6d $(D_{N+2}, D_{N+2})$ minimal conformal matter theory'' 
\cite{ DelZotto:2014hpa} are expected to be realized.
This six-dimensional superconformal field theory is known also as 
the UV fixed point theory for the 6d $\mathcal{N}=(1,0)$ $Sp(N-2)$ gauge theory with $N_f=2N+4$ flavors and one tensor multiplet \cite{Seiberg:1996qx, Danielsson:1997kt}, 
which is realized by a type IIA brane configuration \cite{Hanany:1997gh, Brunner:1997gf}.

In \cite{Gaiotto:2015una}, it is proposed that  
the 5d $\mathcal{N}=1$ $SU(N)$ theory with $N_f$ flavors
with the Chern-Simons level $\kappa=N+2-N_f/2$ and
the 5d $\mathcal{N}=1$ $Sp(N-1)$ theory with $N_f$ flavors 
have an identical UV fixed point based on the discussion of ``duality wall''.
Combining this with the conjectures mentioned above, it is indicated that the three theories,
\begin{itemize}
\item 6d $Sp(N-2)$ theory with $N_f=2N+4$ flavors and a tensor multiplet, 
\item 5d $Sp(N-1)$ theory with $N_f=2N+4$ flavors,
\item 5d $SU(N)$ theory with $N_f=2N+4$ flavors,
\end{itemize}
have the identical 6d UV fixed point of the 6d $(D_{N+2}, D_{N+2})$ minimal conformal matter theory. In other words, the three different descriptions are obtained by turning on different types of deformations from the 6d superconformal field theory. 
This ``equivalence'' has a natural explanation \cite{Hayashi:2015zka, Hayashi:2015fsa} 
in terms of the decomposition of an $O7^-$-plane into two $[p,q]$ 7-branes, namely
a $[1,1]$ 7-brane and a $[1,-1]$ 7-brane \cite{Sen:1996vd}.
The type IIA brane setup for the 6d $Sp(N-2)$ theory includes one $O8^-$-plane.
Compactifying one of the direction and performing T-duality, the $O8^-$-plane is converted into two $O7^-$-planes.
When we decompose one $O7^-$-plane out of the two, we obtain the brane web diagram for the 5d $Sp(N-1)$ theory.
On the other hand, when we decompose both of the two $O7^-$-planes, we obtain the brane web diagram for the 5d $SU(N)$ theory.

We would like to give further evidence to this equivalence
and study it at a more quantitative level.
One of the best playground is the BPS counting.
Indeed in \cite{Kim:2014dza}, it has been checked that the elliptic genus for the E-string theory agrees with
the Nekrasov partition function for the 5d $Sp(1)$ gauge theory with 8 flavors \cite{Hwang:2014uwa}.
We would like to generalize this relation to a higher rank case.
For 6d $Sp(N-2)$ theory with $N_f=2N+4$ flavors, it is straightforward for one to obtain an explicit expression for the elliptic genus \cite{Kim:2014dza, Kim:2015fxa}.
However, 
there had been a difficulty in computing the Nekrasov partition function for the 5d $Sp(N-1)$ gauge theory with $N_f=2N+4$ flavors and also for the 5d $SU(N)$ gauge theory with $N_f=2N+4$ flavors due to a difficulty associated to too many flavors. 
Moreover, we need 
to have a better understanding on the relations among parameters, in addition to the one studied in \cite{Gaiotto:2015una}.

In this paper, we demonstrate that the $(p, q)$ 5-brane web diagram, or the Tao diagram, is useful for overcoming the difficulties. Remarkably, the Tao diagram enables us to compute the 5d $SU(N)$ Nekrasov partition function and also to establish the maps among the parameters,
especially between the 5d $Sp(N-1)$ gauge theory and the 5d $SU(N)$ gauge theory.
For simplicity, we concentrate on the case $N=3$: we check that
\begin{itemize} 
\item the elliptic genus for the 6d $Sp(1)$ gauge theory with $10$ flavors and a tensor multiplet, 
\item the Nekrasov partition function for the 5d $Sp(2)$ gauge theory with $10$ flavors,
\item the Nekrasov partition function for the 5d $SU(3)$ gauge theory with $10$ flavors,
\end{itemize} 
all agree with each other under certain maps among the parameters, which we also determine.
Generalization to an even higher rank case would be straightforward.

Here, we would like to comment a difference between our claim 
and the ``$SU$--$Sp$ duality'' discussed in \cite{Gaiotto:2015una}.
In \cite{Gaiotto:2015una}, it is claimed that the 5d $Sp(N-2)$ Nekrasov partition function
and 5d $SU(N-1)$ Nekrasov partition function are related to each other via ``elliptic Fourier transformation'',
which involves 
a non-trivial integral in terms of the Coulomb branch moduli. 
Instead, our claim is that the 5d $Sp(2)$ Nekrasov partition function
and the 5d $SU(3)$ Nekrasov partition function is simply equal\footnote{More rigorously, it is equal to each other up to ``flop transitions'' for some perturbative factors. 
This point is explained in detail in section \ref{sec:SU3}.}
to each other under a certain map for the gauge theory parameters at least for the unrefined case $\epsilon_1 = - \epsilon_2$.
The maps for the masses and the instanton factor are essentially identical to those in \cite{Gaiotto:2015una}.
In addition to that, we find the map between the Coulomb branch moduli. 
We note that these maps are all obtained intuitively and systematically from the comparison of the corresponding web diagrams.

The organization of this paper is as follows:
In section \ref{sec:maps}, we determine the maps between the gauge theory parameters by using a Higgsing argument as well as 5-brane webs.
In section \ref{sec:SU3}, we compute the Nekrasov partition function of the 5d $SU(3)$ gauge theory with $10$ flavors by applying the topological vertex formalism to the Tao web diagram. Then, we check its order by order expansion in terms of the instanton fugacity and one Coulomb modulus and we find that 5d $SU(3)$ partition function from the Tao diagram agrees with the 6d elliptic genus.
In section \ref{sec:Sp2}, we compare the 5d Nekrasov partition function of the $Sp(2)$ gauge theory with the 6d elliptic genus up to 1-instanton and check that it is also consistent with this ``equivalence'' claim.
We further give some observation to the 2-instanton contribution of the 5d $Sp(2)$ gauge theory with $10$ flavors.
Section \ref{sec:concl} is devoted to conclusion and discussion. We relegate some technical details of the computation to the appendices.

\bigskip

\section{Maps among three theories}
\label{sec:maps}
In this section, we determine the maps among the 6d $Sp(1)$ gauge theory with $10$ flavors and a tensor multiplet, the 5d $Sp(2)$ gauge theory with $10$ flavors, and the 5d $SU(3)$ gauge theory with $10$ flavors. In order to determine all the relations among the three theories, it is enough to determine independent two relations and hence we focus on the map between the 6d $Sp(1)$ gauge theory with $10$ flavors and a tensor multiplet and the 5d $Sp(2)$ gauge theory with $10$ flavors, and also the map between the 5d $Sp(2)$ gauge theory with $10$ flavors and the 5d $SU(3)$ gauge theory with $10$ flavors.

\subsection{Map between the 6d $Sp(1)$ gauge theory and the 5d $Sp(2)$ gauge theory}
\label{sec:6dSp1to5dSp2}
\begin{figure}[t]
\centering
\includegraphics[width=6cm]{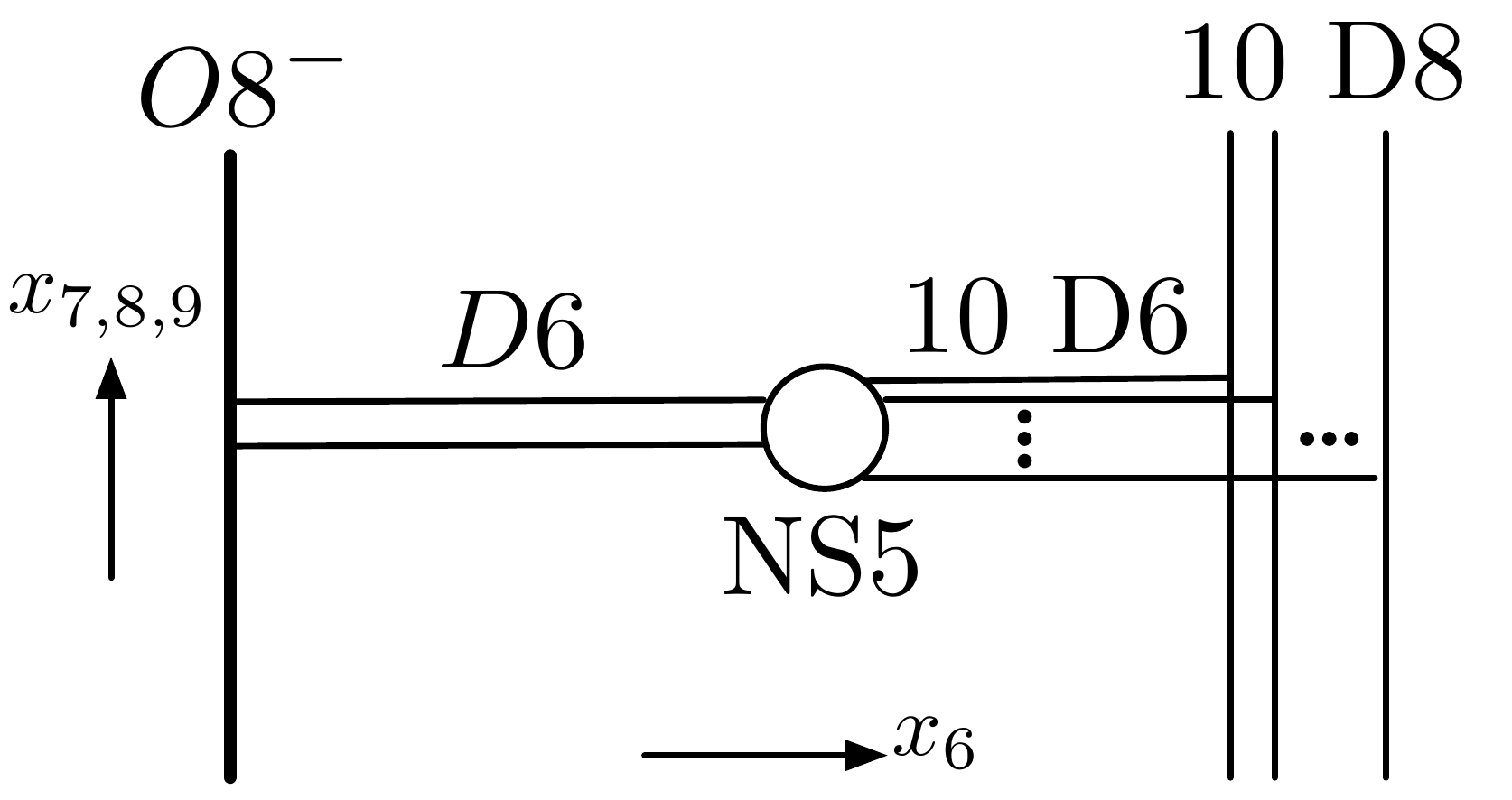}
\caption{The type IIA brane configuration which gives rise to the 6d $Sp(1)$ gauge theory with $10$ flavors and a tensor multiplet.}
\label{fig:6dSp1}
\end{figure}

We first determine the map between the 6d $Sp(1)$ gauge theory with $10$ flavors and a tensor multiplet and the 5d $Sp(2)$ gauge theory with $10$ flavors. The 6d $Sp(1)$ gauge theory with $10$ flavors and a tensor multiplet is realized by a brane configuration in Figure \ref{fig:6dSp1} in type IIA string theory \cite{Hanany:1997gh, Brunner:1997gf}. 
The directions in which the branes extend is summarized in Table \ref{tb:6dbrane}.
\begin{table}[t]
\begin{center}
\begin{tabular}{c|c|cccc|c|c|ccc}
&0&1&2&3&4&5&6&7&8&9\\
\hline
D6&$\times$&$\times$&$\times$&$\times$&$\times$&$\times$&$\times$&&&\\
NS5&$\times$&$\times$&$\times$&$\times$&$\times$&$\times$&&&&\\
D8/O8&$\times$&$\times$&$\times$&$\times$&$\times$&$\times$&&$\times$&$\times$&$\times$\\
\hline
D2&$\times$&&&&&$\times$&$\times$&&&\\
\end{tabular}
\caption{The directions in which branes extend in type IIA string theory.}
\label{tb:6dbrane}
\end{center}
\end{table}
The brane configuration consists of one D6-branes (or two D6-branes including the mirror image), an $O8^-$-plane which gives the $Sp(1)$ gauge symmetry, and $10$ D8-branes which yield the $10$ hypermultiplets in the fundamental representation of the $Sp(1)$. The length between the $O8^-$-plane and the NS5-brane gives a vacuum expectation value (vev) of a scalar field in the tensor multiplet. D2-branes suspended between the $O8^-$-plane and the NS5-brane realize self-dual strings in the 6d theory. For example, $k$ D2-branes mean $k$ strings. 

The elliptic genus of the 6d $Sp(1)$ gauge theory with $10$ flavors or the partition function of a theory on the self-dual strings of the 6d theory wrapping a torus may be written by
\be
\tilde{Z}_{\text{6d Sp(1)}} = 
\tilde{Z}_{(0)} \left( 1 + \sum_{k=1}^{\infty} 
\tilde{Z}_{(k)}\left(g_1, g_2, \tilde{A}, \tilde{y}_1, \cdots, \tilde{y}_{10}, \tilde{q} \right)\phi^k \right), \label{6delliptic}
\ee
where $\tilde{Z}_{(k)}$ stands for the elliptic genus of the $k$-strings while $\tilde{Z}_{(0)}$ is the contribution 
existing even without strings.
$g_1, g_2$ are defined by $g_1 = e^{-\epsilon_1}, g_2=e^{-\epsilon_2}$ where $\epsilon_1, \epsilon_2$ are chemical potentials for the $SO(4)$ global symmetry which rotates the $(x_1, x_2, x_3, x_4)$-plane. In the later calculation, we focus only on a special case where $\epsilon=\epsilon_1 = -\epsilon_2$ with $g=e^{-\epsilon}$. $\tilde{A}$ is the fugacity for the $Sp(1)$ gauge symmetry. 
$\tilde{y}_i\; (i=1, \cdots, 10)$ are  
the fugacities for the $SO(20)$ flavor symmetry. $\tilde{q}$ is given by $\tilde{q} = e^{2\pi i\tau}$ where $\tau$ is the complex structure of the torus. Finally, $\phi$ is the fugacity which counts the number of the self-dual strings.

Let us then move on to the parameters of the 5d $Sp(2)$ gauge theory with $10$ flavors. The 5-brane configuration of the 5d $Sp(2)$ gauge theory with $10$ flavors can be obtained from the type IIA brane configuration in Figure \ref{fig:6dSp1} by first compactifying it on a circle along the $x_5$-direction and then performing T-duality along it \cite{Hayashi:2015zka}. 
\begin{figure}[t]
\centering
\includegraphics[width=8cm]{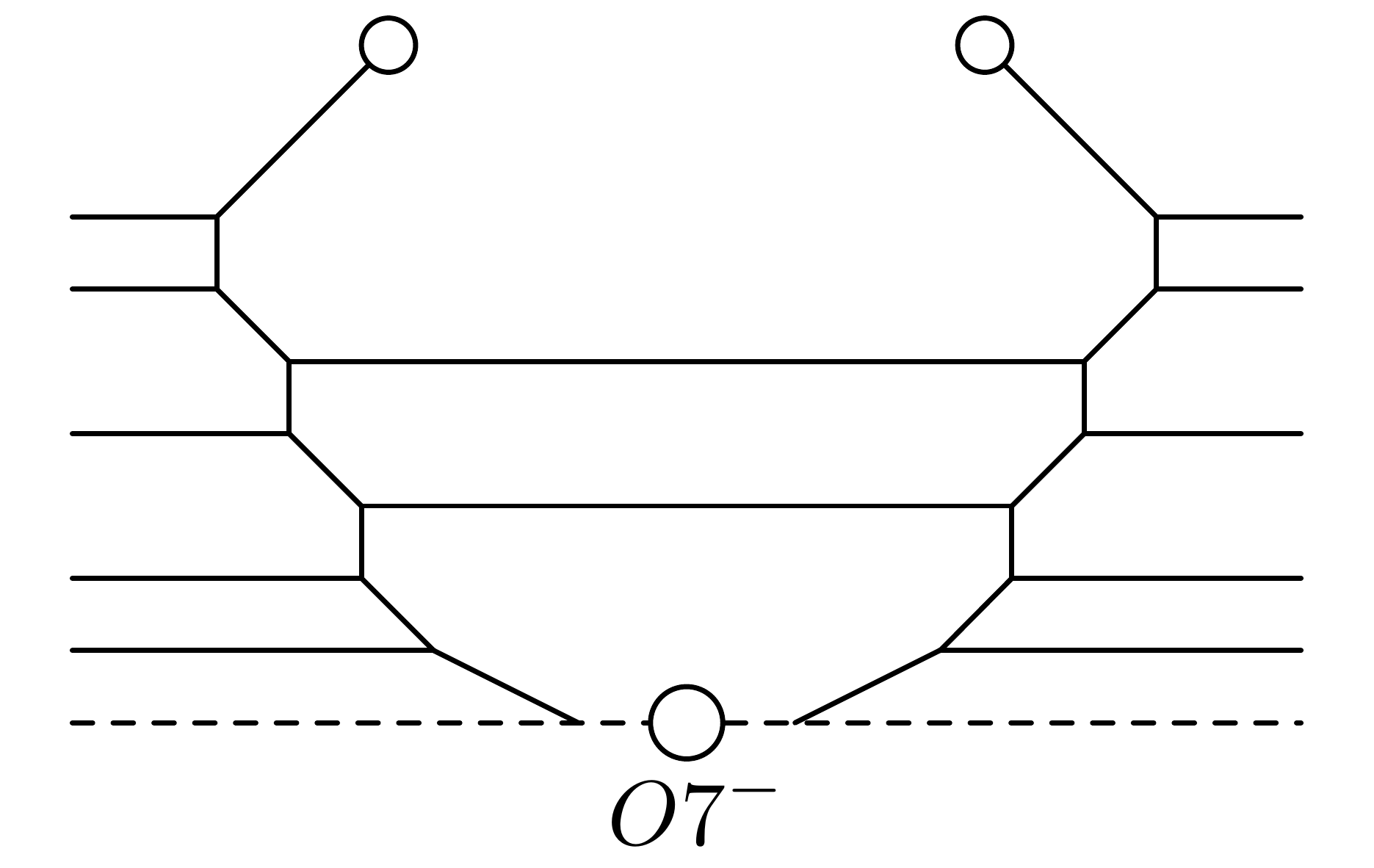}
\caption{The 5-brane web which gives rise to the 5d $Sp(2)$ gauge theory with $10$ flavors. }
\label{fig:5dSp2}
\end{figure}
After the T-duality, the $O8^-$-plane becomes two $O7^-$-planes and the quantum resolution of only one $O7^-$-plane yields a 5-brane web with one remaining $O7^-$-plane whose worldvolume theory is the 5d $Sp(2)$ gauge theory with $10$ flavors. The brane configuration is depicted in Figure \ref{fig:5dSp2} where the horizontal direction is in the $x_6$-direction and the vertical direction is the $x_5$-direction. In the 5-brane web, we have $(p, q)$ 5-branes whose slope in the $(x_6, x_5)$-plane is $\frac{q}{p}$. Furthermore, a $(p, q)$ 7-brane can be put at the end of an external $(p, q)$ 5-brane. The directions in which the branes extend is summarized in Table \ref{tb:5brane}.
\begin{table}[t]
\begin{center}
\begin{tabular}{c|c c c c c | c c | c c c}
& 0 & 1 & 2 & 3 & 4 & 5 & 6 & 7 & 8 & 9\\
\hline
D5 & $\times$ & $\times$ & $\times$ & $\times$ & $\times$ & &$\times$  &&& \\
NS5& $\times$ & $\times$ & $\times$ & $\times$ & $\times$ &$\times$  & &&& \\
$(p, q)$ 5 & $\times$ & $\times$ & $\times$ & $\times$ & $\times$ & \multicolumn{2}{|c|}{\text{angle}}&&& \\
$(p, q)$ 7 & $\times$ & $\times$ & $\times$ & $\times$ & $\times$ &  &  & $\times$ &$\times$  & $\times$
\end{tabular}
\caption{The directions in which 5-branes and 7-branes extend in type IIB string theory.}
\label{tb:5brane}
\end{center}
\end{table}
The two color D5-branes with the $O7^-$-plane yields the $Sp(2)$ gauge symmetry and the $10$ flavor (or external) D5-branes give the $10$ hypermultiplets in the fundamental representation of the $Sp(2)$. 

We consider a Coulomb branch of the 5d $Sp(2)$ gauge theory with $10$ flavors. The theory has two Coulomb branch moduli $a'_1, a'_2$ and ten mass parameters $m'_i$'s $(i=1, \cdots, 10)$ by which we define $A'_1 = e^{-a'_1}, A'_2 = e^{-a'_2}$ and $y'_i = e^{-m'_i}$ for $i=1, \cdots, 10$. The theory also has the instanton fugacity $q'$, which is related to the gauge coupling. In order to calculate the partition function of the 5d theory, we also turn on the $\Omega$-background with the $\Omega$-deformation parameter $\epsilon (=\epsilon_1 = -\epsilon_2)$ from which we define $g=e^{-\epsilon}$. Then, the partition function of the 5d theory may be given by 
\be
Z_{\text{5d Sp(2)}}^{\prime} = Z'_{0} \left( 1 + 
\sum_{k=1}^{\infty} Z'_k\left(g, A'_1, A'_2, y'_1, \cdots, y'_{10}\right) q'^k
\right), \label{5dSp2part}
\ee   
where $Z'_k$ represents the $k$-instanton partition function while $Z'_{0}$ is the perturbative contribution.
Here we used the same $g$ as the $g$ defined by the variables in the elliptic genus of the 6d $Sp(1)$ gauge theory since they can be identified with each other. 

The claim that the UV completion of the 5d $Sp(2)$ gauge theory with $10$ flavors is identical to that of the 6d $Sp(1)$ gauge theory with $10$ flavors and a tensor multiplet means that the elliptic genus \eqref{6delliptic} should be equal to the 5d partition function \eqref{5dSp2part}
\be
\tilde{Z}_{\text{6d Sp(1)}}\left(g, \tilde{A}, \tilde{y}_1, \cdots, \tilde{y}_{10}, \tilde{q} , \phi\right) = Z_{\text{5d Sp(2)}}^{\prime}\left(g, A'_1, A'_2, y'_1, \cdots, y'_{10}, q'\right),
\ee
under certain reparameterization, which we determine here. 

In order to determine the relation between the parameters in the elliptic genus of the 6d $Sp(1)$ gauge theory and those of the 5d $Sp(2)$ gauge theory, we utilize a Higgsing to the E-string theory, which is given by a 6d ``$Sp(0)$'' gauge theory with $8$ flavors and a tensor multiplet on a tensorial Coulomb branch. A circle compactification of the E-string theory with Wilson lines along it gives a 5d $Sp(1)$ gauge theory with $8$ flavors. Therefore the same Higgsing should reduce the 5d $Sp(2)$ gauge theory with $10$ flavors to the 5d $Sp(1)$ gauge theory with $8$ flavors.  
The relations between the parameters in the elliptic genus of the 6d $Sp(0)$ gauge theory 
and the parameters of the 5d $Sp(1)$ gauge theory with $8$ flavors has been known in \cite{Hwang:2014uwa, Kim:2014dza} and hence we can make use of them to deduce the relations between the parameters in the elliptic genus of the 6d $Sp(1)$ gauge theory and the parameters of the 5d $Sp(2)$ gauge theory.

Let us first look into the Higgsing from the 6d $Sp(1)$ gauge theory with $10$ flavors and a tensor multiplet to the 6d $Sp(0)$ gauge theory with $8$ flavors and a tensor multiplet. The explicit expression of one string is given by \cite{Kim:2015fxa}
\begin{align}
\tilde{Z}_{(1)}\left(g, g^{-1}, \tilde{A}, \tilde{y}_1, \cdots, \tilde{y}_{10}, \tilde{q} \right) = -\frac12 \frac{\eta^2}{\theta_{1}(g)\theta_1(g^{-1})}\sum_{I=1}^{4}\frac{\eta^2}{\theta_I(\tilde{A})\theta_I(\tilde{A}^{-1})}\prod_{l=1}^{10}\frac{\theta_I(\tilde{y}_l)}{\eta}.\label{Sp1onestring}
\end{align}
where we have already set $\epsilon=\epsilon_1 = -\epsilon_2$. Then, the Higgsing towards the E-sting elliptic genus may be triggered by setting\footnote{In general, one can choose arbitrary two parameters $\tilde{y}_j, \tilde{y}_k$ with fixed $j, k$ among the ten parameters $\tilde{y}_i \;(i=1, \cdots, 10)$ and set
\begin{align*}
\tilde{A} = \tilde{y}_j = \tilde{y}_k^{-1}.	
\end{align*}
for the Higgsing. In \eqref{6dSp1toEstring}, we choose particular two mass parameters $\tilde{y}_3, \tilde{y}_8$ for concreteness.}
\be
\tilde{A} = \tilde{y}_3 = \tilde{y}_8^{-1},\label{6dSp1toEstring}
\ee
in the case when $\epsilon_1 = -\epsilon_2 = \epsilon$. 
Indeed, inserting \eqref{6dSp1toEstring} to \eqref{Sp1onestring} yields
\begin{align}
-\frac12 \frac{\eta^2}{\theta_{1}(g)\theta_1(g^{-1})}\sum_{I=1}^{4}\prod_{l=1,2,4, \cdots, 7, 9, 10}\frac{\theta_I(\tilde{y}_l)}{\eta},
\end{align}
which exactly agrees with the one string result of the elliptic genus of the E-string if the other parameters are naturally related to the parameters in the elliptic genus of the E-string theory. Namely, $\tilde{y}_i\;(i=1, 2, 4, \cdots 7, 9)$ and $\tilde{y}_{10}$ become the eight fugacities for the $E_8$ flavor symmetry, and $\tilde{q}$ and $\phi$ remain the same. 

Similarly, the Higgsing from the 5d $Sp(2)$ gauge theory with $10$ flavors to the 5d $Sp(1)$ gauge theory with $8$ flavors can be also deduced from the explicit expressions of their partition functions of the 5d $Sp$ gauge theories. In fact, the Higgsing may be realized by essentially the same condition 
\be
A'_2 = y'_3 = y_8^{\prime -1}, \label{5dSp2toSp1}
\ee
in the case when $\epsilon_1 = -\epsilon_2 = \epsilon$. We again chose the particular two masses $m'_3, m'_8$ among the ten masses for concreteness but we can in general choose any two masses for the Higgsing. The other parameters are again naturally related to the parameters of the 5d $Sp(1)$ gauge theory with $8$ flavors, namely $a'_1$ becomes the Coulomb branch modulus of the 5d $Sp(1)$ gauge theory, $m'_i\;(i=1, 2, 4, \cdots, 7, 9)$ and $m'_{10}$ are the eight mass parameters for the $8$ flavors, $\epsilon$ is the $\Omega$-deformation parameter and $q'$ is the instanton fugacity of the 5d $Sp(1)$ gauge theory. 

Then, the relation between the parameters in the elliptic genus of the E-string and the parameters of the 5d $Sp(1)$ gauge theory with $8$ flavors is known in \cite{Hwang:2014uwa, Kim:2014dza} and non-trivial relations are
\be
\tilde{y}_{10}=y^{\prime }_{10}q^{\prime -2}, \qquad \phi = A'_1q' y^{\prime -1}_{10}, \qquad \tilde{q} =q^{\prime 2}.\label{Estring.relation}
\ee
The other relations are trivial and given by 
$\tilde{y}_i = y'_i\; (i=1, 2, 4, 5, 6, 7, 9)$ together with the same $g$. Furthermore, the two Higgsings \eqref{6dSp1toEstring} and \eqref{5dSp2toSp1} are the same Higgsing and this suggests that they can be identified with each other, namely
\be
\tilde{A} = A'_2, \quad \tilde{y}_3 = y'_3, \quad \tilde{y}_8 = y'_8. \label{otherrelation}
\ee

Therefore, we propose that the relations between the parameters of the elliptic genus of the 6d $Sp(1)$ gauge theory and the parameters of the 5d $Sp(2)$ gauge theory are 
\bea
\tilde{y}_{10}&=&y^{\prime }_{10}q^{\prime -2}, \label{6dSp1to5dSp2.map1}\\
\phi &=& A'_1q' y^{\prime -1}_{10}, \label{6dSp1to5dSp2.map2}
\eea
and also
\be
\tilde{q} = q^{\prime 2}, \qquad \tilde{A} = A_2', \qquad \tilde{y}_i = y'_i,\;(i=1, \cdots, 9).\label{6dSp1to5dSp2.map3}
\ee
The relation \eqref{6dSp1to5dSp2.map1} can be also written as
\be
y'_{10} = \tilde{y}_{10}\tilde{q}. \label{wilsonlineshift}
\ee
When we denote the radius of the compactification circle from 6d to 5d by $R$, $\tilde{q}$ may be roughly given by $\tilde{q} \sim e^{-\frac{1}{R}}$. A Wilson line along the circle is given by $A \sim \frac{1}{R}$ and hence $\tilde{q} \sim e^{-A}$. Since the Wilson line in the type IIA picture corresponds to the position of the D7-branes along the compactified circle in the type IIB picture, the relation \eqref{wilsonlineshift} means that the 10th D7-brane goes around the circle and goes back to the original point. Thus, even after shifting the position of the 10th D7-brane, we can still put the $10$ D7-branes at the position of one $O7^-$-plane, indicating that the theory still has the $SO(20)$ global symmetry.\footnote{We note that the $SO(20)$ flavor symmetry of the 5d theories is different from the $SO(20)$ flavor symmetry of the 6d theory. When we compactify a 6d theory on a circle, we introduce a Wilson line which breaks 6d flavor symmetry. Here, the Wilson line that we introduced produces the 5d $SO(20)$ flavor symmetry.}
This is indeed necessary to reproduce the expected global symmetry for the 5d $Sp(2)$ gauge theory. 

On the other hand, the combination of \eqref{6dSp1to5dSp2.map2} may look more involved. We here provide further support for the relation \eqref{6dSp1to5dSp2.map2}. In order to see it, we analyze the invariance of the elliptic genus of the 6d $Sp(1) $ gauge theory under a Weyl transformation of the 5d $SO(20)$ flavor symmetry. Although the elliptic genus of the 6d $Sp(1)$ gauge theory is manifestly invariant under a Weyl transformation of the 6d $SO(20)$ flavor symmetry, the invariance under a Weyl transformation of the 5d $SO(20)$ flavor symmetry cannot not be seen manifestly and hence it induces a non-trivial transformation rule for the parameters in the elliptic genus. 

For example, let us see the invariance of the elliptic genus of one string \eqref{Sp1onestring} under the exchange between $y'_{10} \leftrightarrow y'_9$, which is a Weyl transformation of the 5d $SO(20)$ flavor symmetry. Assuming \eqref{6dSp1to5dSp2.map2}, the Weyl transformation induces the map
\bea
\tilde{y}_{10} \rightarrow \tilde{q}^{-1}\tilde{y}_9, \qquad \tilde{y}_9 \rightarrow \tilde{q} \tilde{y}_{10}. \label{Weyltransform1}
\eea
Inserting the transformation \eqref{Weyltransform1} into \eqref{Sp1onestring} yields
\be
\tilde{Z}_{(1)} \rightarrow  \left[-\frac12 \frac{\eta^2}{\theta_{1}(g)\theta_1(g^{-1})}\sum_{I=1}^{4}\frac{\eta^2}{\theta_I(\tilde{A})\theta_I(\tilde{A}^{-1})}\prod_{l=1}^{10}\frac{\theta_I(\tilde{y}_l)}{\eta}\right] \times \tilde{q}^{-1}\tilde{y}_9\tilde{y}_{10}^{-1}. \label{onestringWeyl}
\ee
Since the elliptic genus of one string, {\it i.e.} $\tilde{Z}_{(1)}\phi$, should be also invariant under the Weyl transformation, the transformation rule for the $\phi$ turns out to be
\be
\phi \rightarrow \tilde{q} \tilde{y}_9^{-1}\tilde{y}_{10}\phi\label{Weyltransform2}
\ee
so that it cancels the extra factor in \eqref{onestringWeyl}. 

Note here that the Coulomb branch modulus $A'_1$ should be invariant under the Weyl transformation and hence the combination $\phi \tilde{q}^{\frac{1}{2}} \tilde{y}_{10}$ should be also invariant under the exchange. In fact, the Weyl transformation \eqref{Weyltransform1} and \eqref{Weyltransform2}, which is induced by the exchange between $y'_{10}$ and $y'_9$, acts on the combination $\phi \tilde{q}^{\frac{1}{2}} \tilde{y}_{10}$ as
\be
A'_1 = \phi \tilde{q}^{\frac{1}{2}} \tilde{y}_{10} \rightarrow \left(\tilde{q}　\tilde{y}_9^{-1}\tilde{y}_{10} \phi\right)\tilde{q}^{\frac{1}{2}} \left(\tilde{q}^{-1} \tilde{y}_{9}\right) = \phi \tilde{q}^{\frac{1}{2}} \tilde{y}_{10} = A'_1. 
\ee
Therefore, the combination $\phi \tilde{q}^{\frac{1}{2}} \tilde{y}_{10}$ is invariant under the Weyl transformation $y'_9 \leftrightarrow y'_{10}$ and this is consistent with the fact that it corresponds to the 5d Coulomb branch modulus $A'_1$, which should be invariant under the Weyl transformation of the 5d $SO(20)$ flavor symmetry. 
Analogous discussion on the invariance of Coulomb moduli parameter 
under Weyl transformation is also in \cite{Mitev:2014jza, Kim:2015jba}.

\subsection{Map between the 5d $Sp(2)$ gauge theory and the 5d $SU(3)$ gauge theory}\label{sec:sp2su3map}
We then turn to the map between the 5d $Sp(2)$ gauge theory with $10$ flavors and the 5d $SU(3)$ gauge theory with $10$ flavors. A part of the map is found in \cite{Gaiotto:2015una} and it was determined experimentally from the requirement that the hemisphere index of the 5d $Sp(N-1)$ gauge theory with $N_f$ flavors is equal to the hemisphere index of the 5d $SU(N)$ gauge theory with $N_f$ flavors after performing the duality map. We here derive the map from the 5-brane webs for the two theories. Furthermore, we will find the map of the parameters between the partition functions of the 5d theories and hence the map includes the one for the Coulomb branch moduli, which is not considered in \cite{Gaiotto:2015una}.   

\subsubsection{Parameterization of the 5d $SU(3)$ gauge theory with $10$ flavors}
\begin{figure}[t]
\centering
\includegraphics[width=6cm]{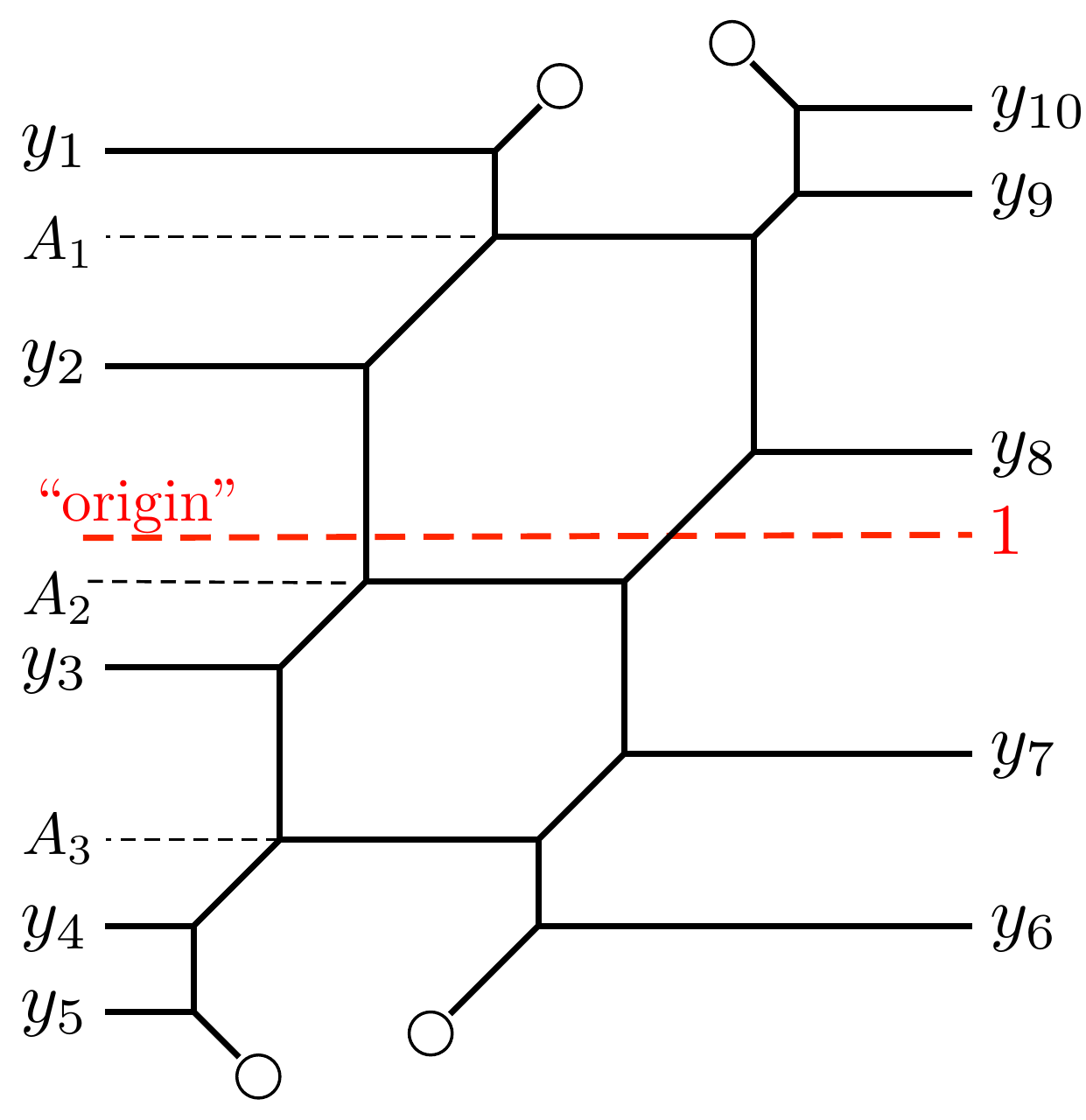}
\caption{The 5-brane web which gives rise to the 5d $SU(3)$ gauge theory with $10$ flavors.}
\label{fig:5dSU3}
\end{figure}

In order to see the relation between the parameters of the 5d $Sp(2)$ gauge theory with $10$ flavors and those of the 5d $SU(3)$ gauge theory with $10$ flavors from the 5-brane webs, we first determine how the parameters of the 5d $SU(3)$ gauge theory with $10$ flavors are related to the parameters of the corresponding 5-brane web. The 5-brane web for the 5d $SU(3)$ gauge theory with $10$ flavors can be obtained by further resolving the remaining $O7^-$-plane in Figure \ref{fig:5dSp2}. The resolution of the $O7^-$-plane and also some changes\footnote{This process is equivalent to flop transitions in the dual M-theory geometry.} of the parameters in the web yield a 5-brane web in Figure \ref{fig:5dSU3}. 
From the brane configuration in Figure \ref{fig:6dSp1}, the 5-brane web in Figure \ref{fig:5dSU3} can be obtained by a circle compactification and T-duality along the $x_5$-direction, 
and then resolving the resulting two $O7^-$-planes. The three color (or internal) D5-branes yield the $SU(3)$ gauge symmetry and the $10$ flavor (or external) D5-branes give the $10$ hypermultiplets in the fundamental representation of the $SU(3)$.

The 5d $SU(3)$ gauge theory have three (but two independent) Coulomb branch moduli $a_1, a_2, a_3$ which satisfy $a_1+a_2+a_3=0$, ten masses $m_i$ $(i=1, \cdots, 10)$ for the $10$ flavors and the instanton fugacity $q$. All the moduli and the parameters are related to the lengths in the 5-brane in Figure \ref{fig:5dSU3}. The three Coulomb branch moduli are related to the position in the vertical direction of the three color D5-branes and we define
\be
A_1 = e^{-a_1}, \quad A_2 = e^{-a_2}, \quad A_3 = e^{-a_3}. \label{CoulombSU3}
\ee
Due to the traceless condition, not all the $A_1, A_2, A_3$ are independent but satisfy the condition
\be
A_1A_2A_3=1. \label{traceless}
\ee
This relation defines the ``origin'' in the vertical direction. Namely, the vertical position is always measured from the origin defined by \eqref{traceless} for determining the parameters of the 5d $SU(3)$ gauge theory with $10$ flavors. On the other hand, the mass parameters for the $10$ hypermultiplets are given by the vertical position of the flavor D5-branes and we define
\be
y_i = e^{-m_i}, \quad i=1, \cdots, 10. \label{eq:yi}
\ee
The explicit locations of the $A_1, A_2, A_3$ and $y_i, \; (i=1, \cdots , 10)$ are depicted in Figure \ref{fig:5dSU3}. 

\begin{figure}[t]
\centering
\includegraphics[width=10cm]{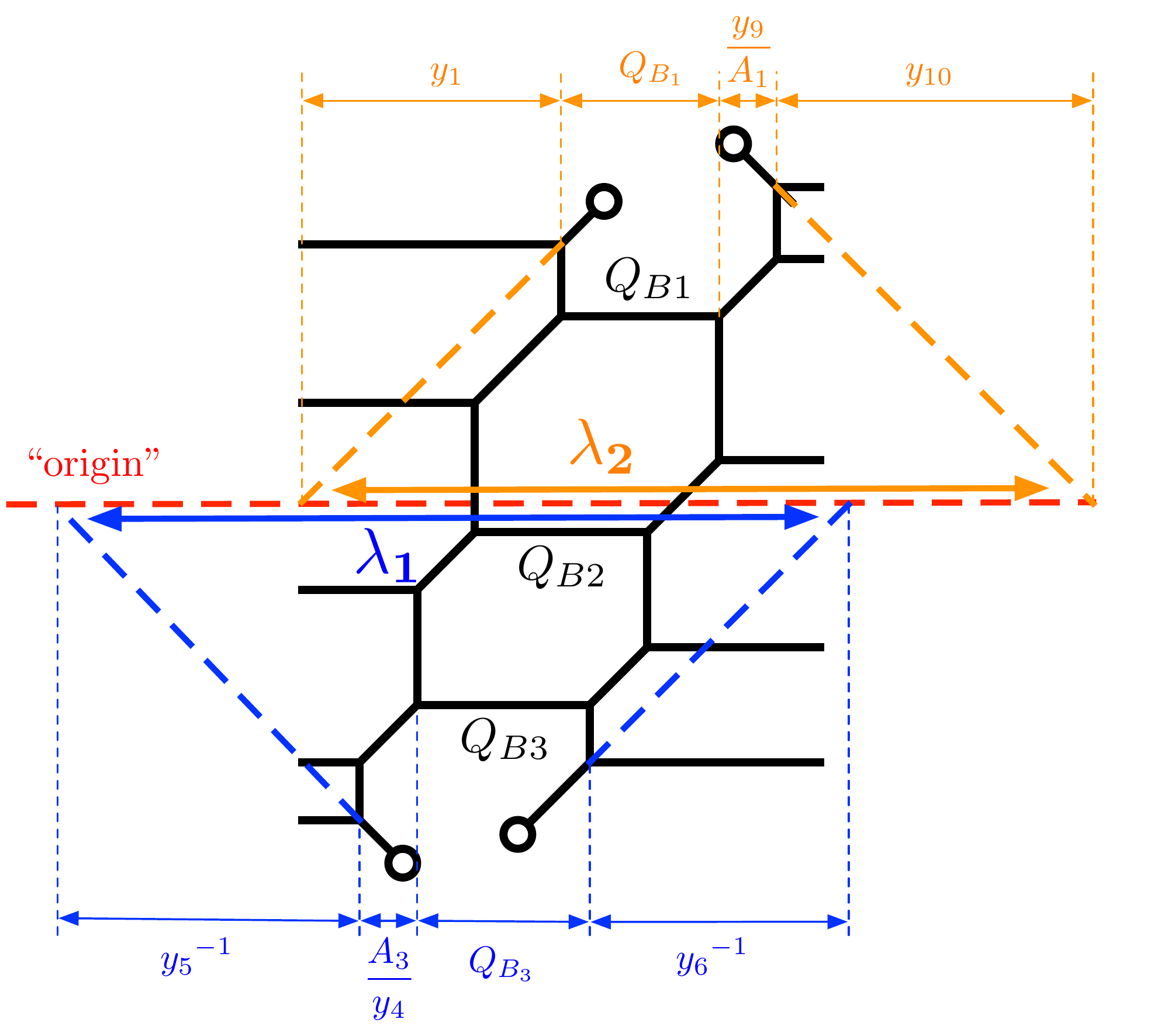}
\caption{The extrapolation of the upper and lower external 5-branes to the origin. }
\label{fig:SU3instanton}
\end{figure}

The last parameter we need to determine is the instanton fugacity of the 5d $SU(3)$ gauge theory. The instanton fugacity is related to the exponential of (the minus of) the average of the length $L_1$ which is obtained by extrapolating the two lower external 5-branes to the origin and the length $L_2$ which is obtained by extrapolating the two upper external 5-branes to the origin \cite{Bao:2011rc}. The schematic picture is depicted in Figure \ref{fig:SU3instanton}.
The explicit expression for the length $L_1, L_2$ are given by
\bea
\lambda_1 &=& e^{-L_1} = y_5^{-1}\left(A_3y_4^{-1}\right)Q_{B_3}y_6^{-1}, \label{lambda1}\\
\lambda_2 &=& e^{-L_2} = y_1Q_{B_1}\left(y_9A_1^{-1}\right)y_{10},  \label{lambda2}
\eea
where $Q_{B_1}, Q_{B_3}$ are $Q_{B_1}=e^{-l_1}, Q_{B_3} = e^{-l_3}$ with $l_1=[\text{The length of the top color D5-brane}]$ and $l_3=[\text{The length of the lowest color D5-brane}]$\footnote{Similarly we define $Q_{B_2} = e^{-l_2}$ with $l_2=[\text{The length of the middle color D5-brane}]$ as in Figure \ref{fig:SU3instanton}.}. Therefore, the instanton fugacity $q$ of the 5d $SU(3)$ gauge theory is given by 
\bea
q = \sqrt{\lambda_1\lambda_2} &=&\sqrt{Q_{B_1}Q_{B_3}A_1^{-1}A_3y_1y_4^{-1}y_5^{-1}y_6^{-1}y_9y_{10}}\\
&=& Q_{B_3}A_1^{-1}A_2^{-1}\sqrt{y_1y_2y_3y_4^{-1}y_5^{-1}y_6^{-1}y_7y_8y_9y_{10}}. \label{SU3instanton}
\eea

\subsubsection{The duality relation}

We then identify the parameters of the 5d $Sp(2)$ gauge theory with $10$ flavors from the web diagram in Figure \ref{fig:5dSU3}. The gauge theory parameters are the two Coulomb branch moduli $a'_1, a'_2$, the ten masses $m'_i, (i=1, \cdots, 10)$ of the flavors and the instanton fugacity $q'$. Again, the gauge theory parameters can be identified with some lengths in a 5-brane web in the presence of an $O7^-$-plane. From the 5-brane web in Figure \ref{fig:5dSp2}, the two Coulomb branch moduli are the heights of the two color D5-branes and the ten mass parameters are the heights of the ten flavor D5-branes. However, an important difference from the case of the 5d $SU(3)$ gauge theory is that the parameters and the moduli of the 5d $Sp(2)$ gauge theory are the heights measured from the vertical position of the $O7^-$-plane. Therefore, it is important to determine the position of the $O7^-$-plane before the splitting in the 5-brane web for the 5d $SU(3)$ gauge theory with $10$ flavors in Figure \ref{fig:5dSU3}.

In order to identify the location of the $O7^-$-plane, one needs to closely look at the resolution of an $O7^-$-plane in a simple example of the pure $Sp(1)$ gauge theory. 
\begin{figure}[t]
\centering
\includegraphics[width=6cm]{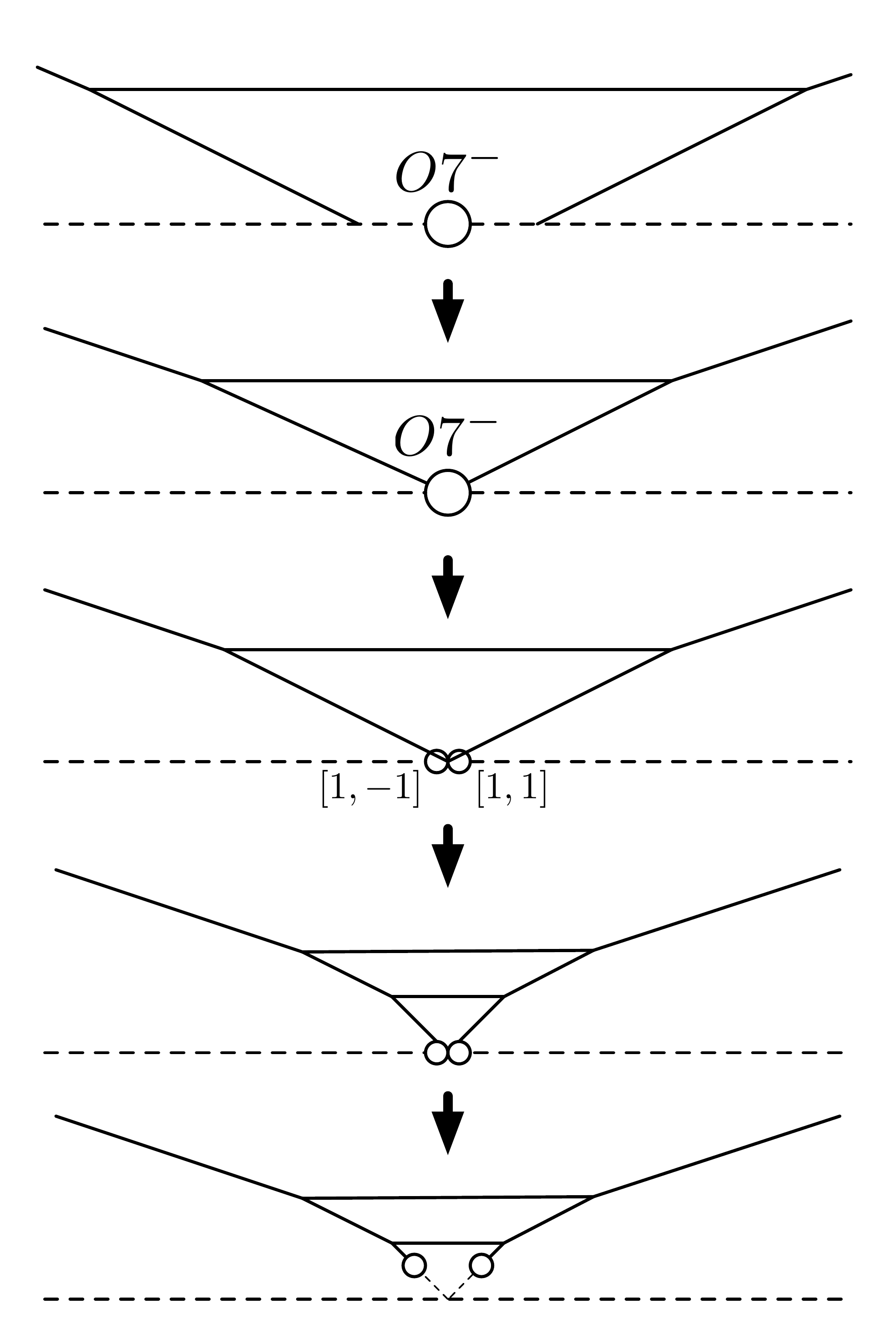}
\caption{Shrinking the Coulomb branch modulus and an effect of the resolution of an $O7^-$-plane. }
\label{fig:O7transition}
\end{figure}
The top figure in Figure \ref{fig:O7transition} stands for a 5-brane web of the 5d pure $Sp(1)$ gauge theory. Let us then think about shrinking the Coulomb branch modulus of the $Sp(1)$ gauge theory which corresponds to shrinking the middle face. Then we reach a configuration in the second figure of Figure \ref{fig:O7transition},
where the middle face reaches to the position of the $O7^-$-plane.
When we further try to shrink the Coulomb branch modulus, 
it is natural to move from $Sp$-like picture to $SU$-like picture by 
 resolving $O7^-$-plane into a $[1, -1]$ 7-brane and a $[1, 1]$ 7-brane at this stage 
 as depicted in the third of Figure \ref{fig:O7transition}.
By further shrinking the Coulomb branch modulus, we now see
that $[1, -1]$ 7-brane and a $[1, 1]$ 7-brane are attached to the 
$(1, -1)$ 5-brane and a $(1, 1)$ 5-brane, respectively,  as in the fourth of Figure \ref{fig:O7transition}.
By shrinking the lengths of the 5-branes which end on the $[1, -1]$ 7-brane and the $[1, 1]$ 7-brane, one finally obtains the lowest figure in Figure \ref{fig:O7transition}, yielding a 5-brane for the pure $SU(2)$ gauge theory. From this figure, it is clear that the location of the original $O7^-$-plane is given by the intersection of the lines extrapolated from the two lower external 5-branes as in the lowest figure in Figure \ref{fig:O7transition}. 

\begin{figure}[t]
\centering
\includegraphics[width=8cm]{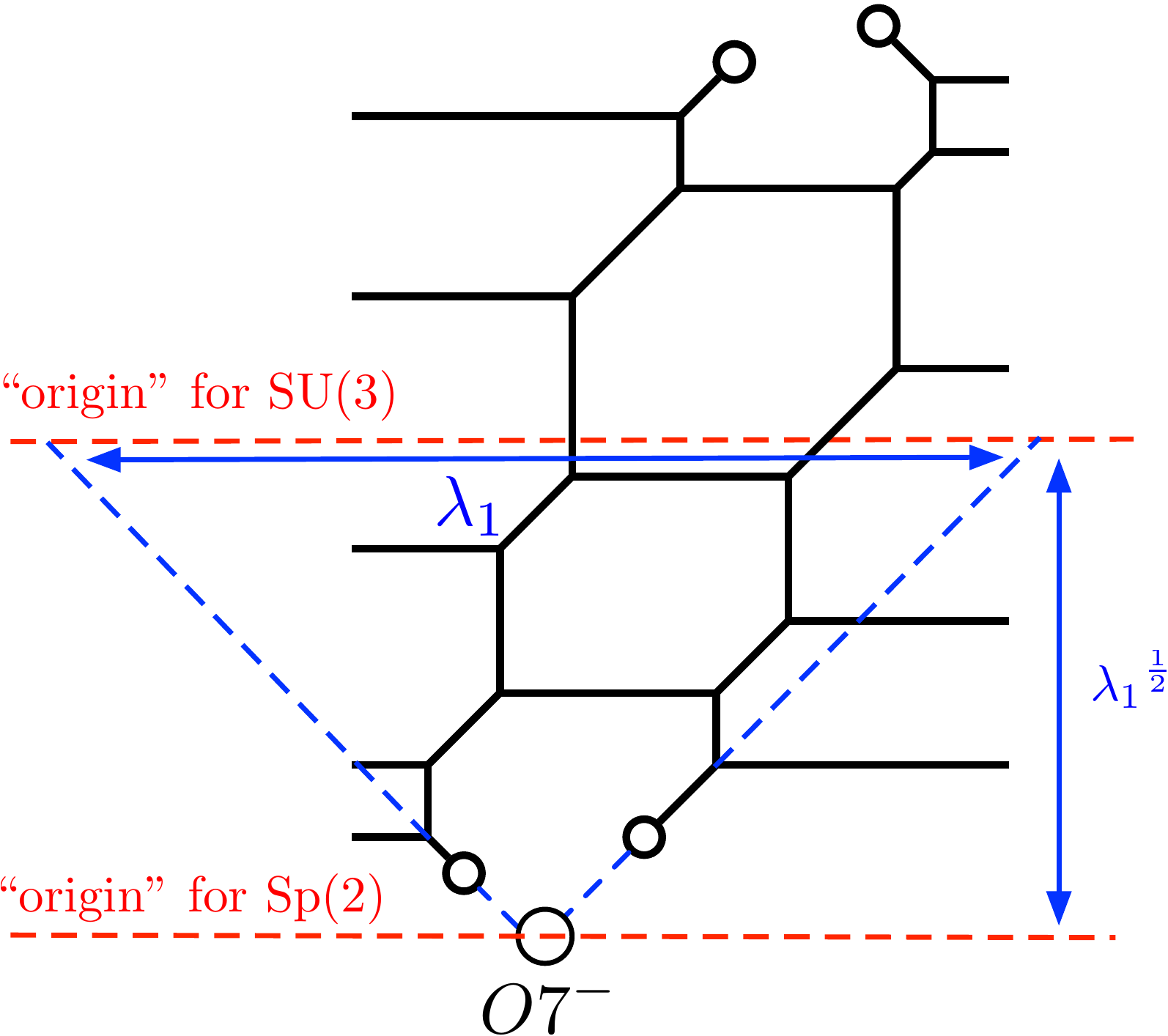}
\caption{The difference of the origin in the vertical direction between the 5-brane webs of the 5d $SU(3)$ gauge theory with $10$ flavors and the 5d $Sp(2)$ gauge theory with $10$ flavors.}
\label{fig:SU3Sp2origin}
\end{figure}

We then apply the procedure of identifying the location of an $O7^-$-plane to the 5-brane web of the 5d $SU(3)$ gauge theory with $10$ flavors in Figure \ref{fig:5dSU3}. As in the case of the pure 5d $SU(2)$ gauge theory, the 7-branes which originate from the $O7^-$-plane in Figure \ref{fig:5dSp2} are the the lowest $[1, -1]$ 7-brane and the lowest $[1, 1]$ 7-brane in Figure \ref{fig:5dSU3}. Hence, the location of the original $O7^-$-plane can be identified by extrapolating the lower external $(1, -1)$ 5-brane and the lower external $(1, 1)$ 5-brane in the lower direction as in Figure \ref{fig:SU3Sp2origin}. 
From the geometry of the 5-brane web in Figure \ref{fig:SU3Sp2origin}, the difference between the origin for the parameters of the 5d $SU(3)$ gauge theory with $10$ flavors and the origin for the parameters of the 5d $Sp(2)$ gauge theory with $10$ flavors is given by
\be
\lambda^{\frac{1}{2}}_1 = q^{\frac{1}{2}}\prod_{i=1}^{10}y_i^{-\frac{1}{4}}, \label{origindiff}
\ee
where we used \eqref{lambda1} and \eqref{SU3instanton}. 

\begin{figure}[t]
\centering
\includegraphics[width=5cm]{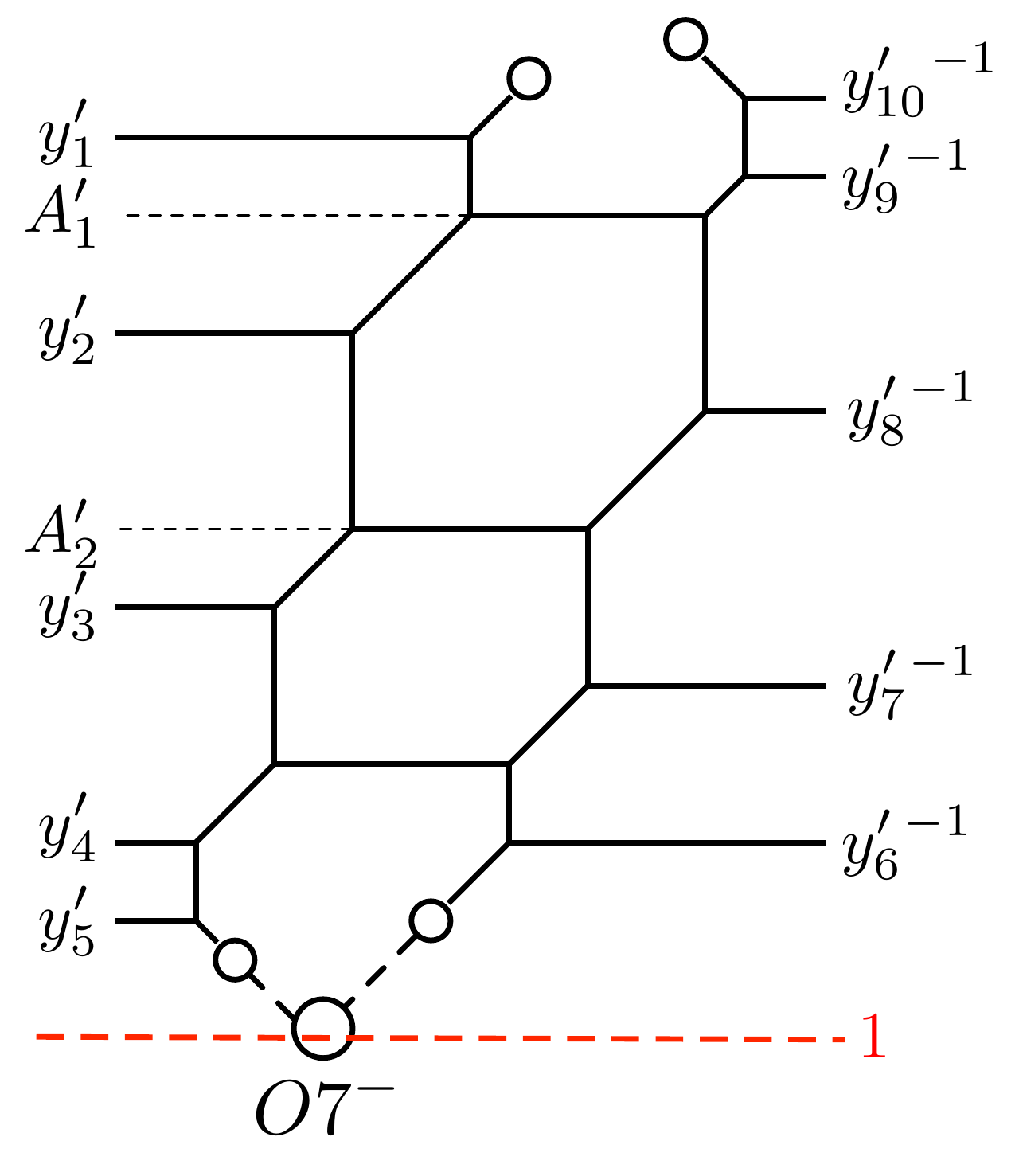}
\caption{The Coulomb branch moduli and the mass parameters of the 5d $Sp(2)$ gauge theory with $10$ flavors in terms of the 5-brane web for the 5d $SU(3)$ gauge theory with $10$ flavors.}
\label{fig:5dSU3Sp2}
\end{figure}

By using the difference of the origin \eqref{origindiff}, it is possible to determine the parameters of the 5d $Sp(2)$ gauge theory with $10$ flavors from the 5-brane web for the 5d $SU(3)$ gauge theory with $10$ flavors as in Figure \ref{fig:5dSU3Sp2}. 
Namely, the Coulomb branch moduli $a'_1, a'_2$ are vertical positions of the top and the second top color D5-branes\footnote{The lowest color D5-brane may be thought of as the one created after the resolution of the $O7^-$-plane. Therefore, we identify the vertical positions of the top and the second top color D5-brane in Figure \ref{fig:5dSU3Sp2} as the two Coulomb branch moduli of the 5d $Sp(2)$ gauge theory.} measured from the location of the $O7^-$-plane.
The difference between \eqref{CoulombSU3} and $A'_1=e^{-a'_1}, A'_2=e^{-a'_2}$ is the difference of the origin \eqref{origindiff}, and hence we obtain
\bea
A'_i &=& \lambda_1^{\frac{1}{2}}A_i =  \left(q^{\frac{1}{2}}\prod_{i=1}^{10}y_i^{-\frac{1}{4}}\right)A_i,\label{Coulombmap}
\eea
for $i=1, 2$.

Analogously, the mass parameters of the ten flavors are related to the heights of the flavor D5-branes measured from the vertical position of the $O7^-$-plane. We use the convention that the mass parameters associated to the heights for the flavor D5-branes on the righthand side have extra minus signs compared to the mass parameters associated to the heights for the flavor D5-branes on the lefthand side. Therefore, we parameterize $y_i=e^{-m_i} \; (i=1, \cdots, 5)$ for the vertical positions of the flavor D5-branes on the left whereas we parameterize $y_i^{-1} = e^{m_i}\; (i=6, \cdots, 10)$ for the vertical positions of the flavor D5-brane on the right as in Figure \ref{fig:5dSU3Sp2}. 

The definition is motivated from the fact that the Higgsing from the 5d $Sp(2)$ gauge theory with $10$ flavors and the zero discrete theta angle\footnote{The discrete theta angle of the 5d $Sp(N)$ gauge theory is not a physical difference when the gauge theory has flavors \cite{Intriligator:1997pq} since it can be absorbed by a sign of a mass parameter. However, we keep using this terminology in this ten flavors case for indicating that two 5d $Sp$ gauge theories with the same discrete theta angle have the same convention for the signs of the mass parameters.} to the 5d $Sp(1)$ gauge theory with $8$ flavors and the zero discrete theta angle is achieved for example by setting $A'_2 = y'_3 = y^{\prime -1}_8$ like \eqref{5dSp2toSp1} as discussed in section \ref{sec:6dSp1to5dSp2}. In terms of the 5-brane web in Figure \ref{fig:5dSU3Sp2}, the Higgsing corresponds to aligning the vertical position of the third flavor D5-brane from the top on the lefthand side,  
the vertical position of the color D5-brane associated to $A'_2$ and the vertical position of the third flavor D5-brane from the top on the righthand side. Therefore, we should associate $y_3$ to the position of the third flavor D5-brane from the top on the left, whereas $y_8^{-1}$ to the position of the third flavor D5-brane from the top on the right. In other words, $-m'_8$ is the height of the flavor D5-brane associated to $y^{\prime -1}_8$ in Figure \ref{fig:5dSU3Sp2}. The same argument essentially holds for the parameters $y'_6, y'_7, y'_8, y'_9, y'_{10}$ and hence the vertical positions of the flavor D5-branes on the righthand side of Figure \ref{fig:5dSU3Sp2} are parametrized by $y^{\prime -1}_6, y^{\prime -1}_7, y^{\prime -1}_8, y^{\prime -1}_9$ and $y^{\prime -1}_{10}$. With this in mind, the difference between $y_i$ and $y^{\prime -1}_i$ for $i=6, \cdots, 10$ as well as the difference between $y_i$ and $y'_i$ for $i=1, \cdots, 5$ are again the difference between the origin \eqref{origindiff} and hence we obtain
\be
y'_i = \lambda_1^{\frac{1}{2}}y_i = \left(q^{\frac{1}{2}}\prod_{i=1}^{10}y_i^{-\frac{1}{4}}\right)y_i \label{massmap1}
\ee
for $i=1, \cdots, 5$ and 
\be
y_i^{\prime -1} = \lambda_1^{\frac{1}{2}}y_i = \left(q^{\frac{1}{2}}\prod_{i=1}^{10}y_i^{-\frac{1}{4}}\right)y_i \label{massmap2}
\ee
for $i=6, \cdots, 10$.

\begin{figure}[t]
\centering
\includegraphics[width=12cm]{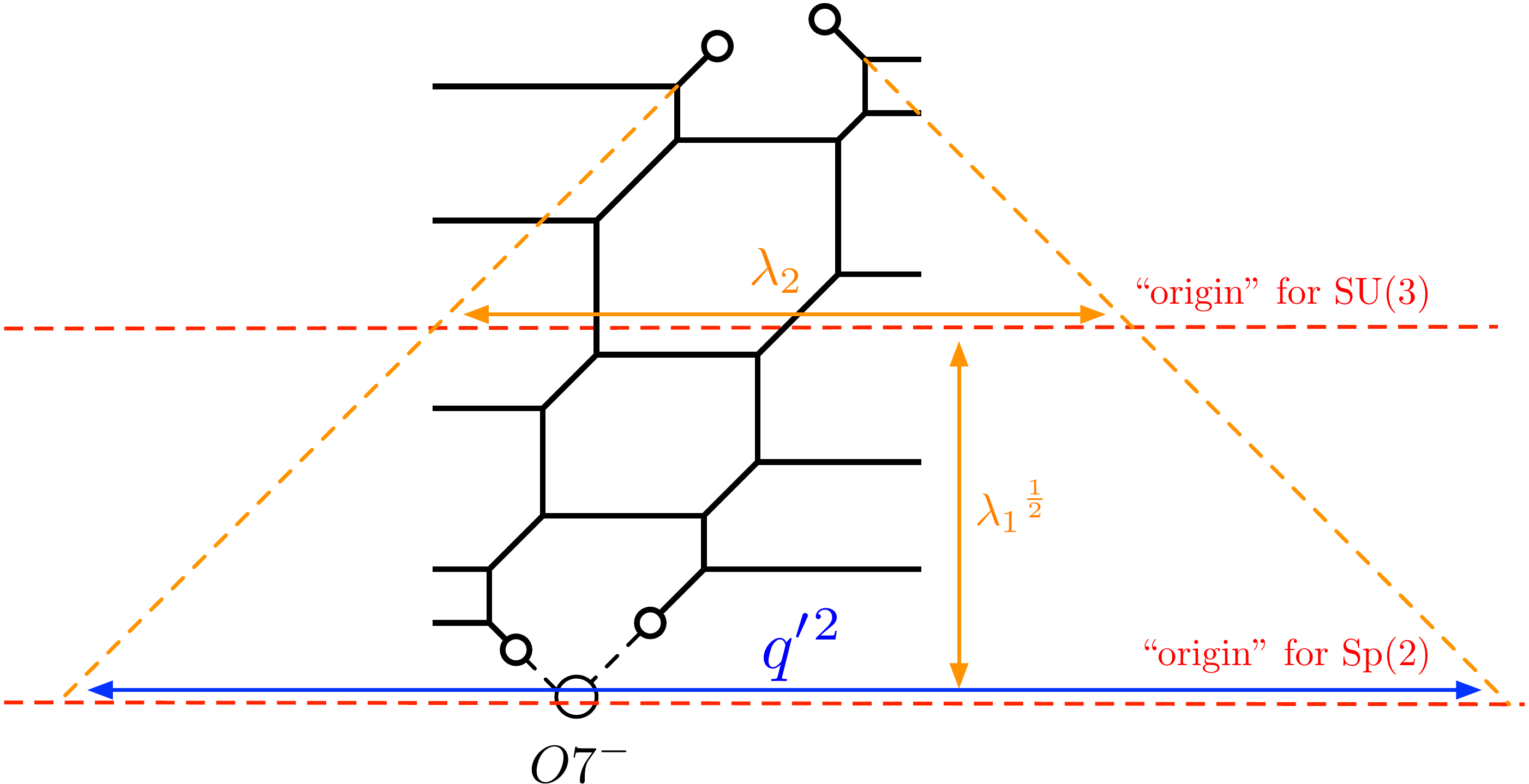}
\caption{The diagramatic identification of the instanton fugacity $q'$ of the 5d $Sp(2)$ gauge theory with $10$ flavors. }
\label{fig:Sp2instanton}
\end{figure}

Let us finally see the relation between the instanton fugacity $q$ of the 5d $SU(3)$ gauge theory with $10$ flavors and the instanton fugacity $q'$ of the 5d $Sp(2)$ gauge theory with $10$ flavors. The square of the instanton fugacity of $q'$ is related to the length between two lines extrapolated from the upper external 5-branes to the origin specified by the vertical position of the $O7^-$-plane. The schematic picture is depicted in Figure \ref{fig:Sp2instanton}. 
The geometry of the 5-brane web in Figure \ref{fig:Sp2instanton} gives
\be
q' = \sqrt{\lambda^{\frac{1}{2}}_1\lambda_2\lambda_1^{\frac{1}{2}}} = q.\label{instantonmap}
\ee
Namely, the instanton fugacity of the 5d $SU(3)$ gauge theory with $10$ flavors is the same as the instanton fugacity of the 5d $Sp(2)$ gauge theory with $10$ flavors.

\bigskip

\section{5d $SU(3)$ gauge theory with $10$ flavors}
\label{sec:SU3}
\begin{figure}[t]
\centering
\begin{tabular}{cc}
\begin{minipage}{0.5\hsize}
\begin{center}
\includegraphics[width=6cm]{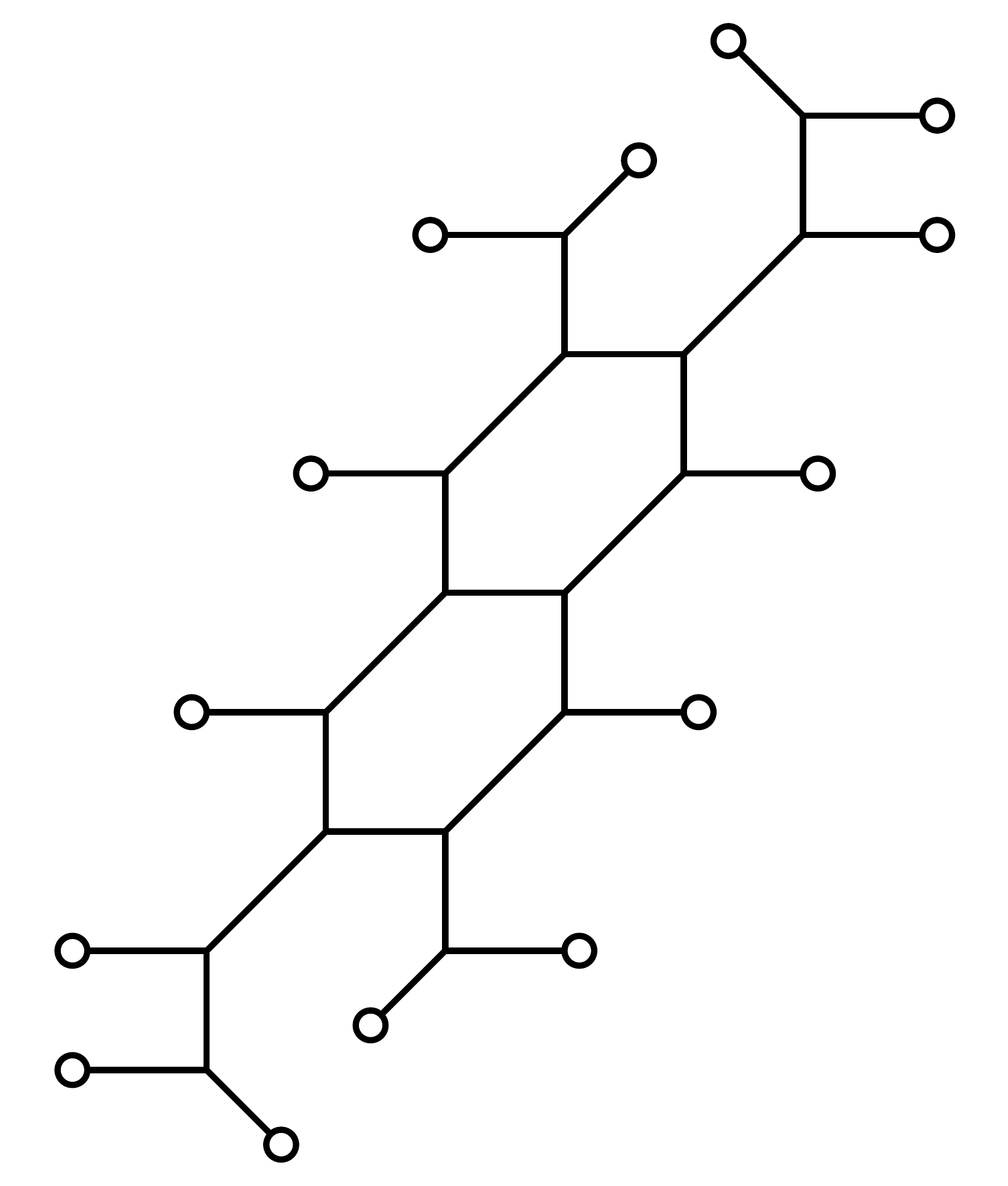}
\end{center}
\end{minipage}
\begin{minipage}{0.5\hsize}
\begin{center}
\includegraphics[width=6cm]{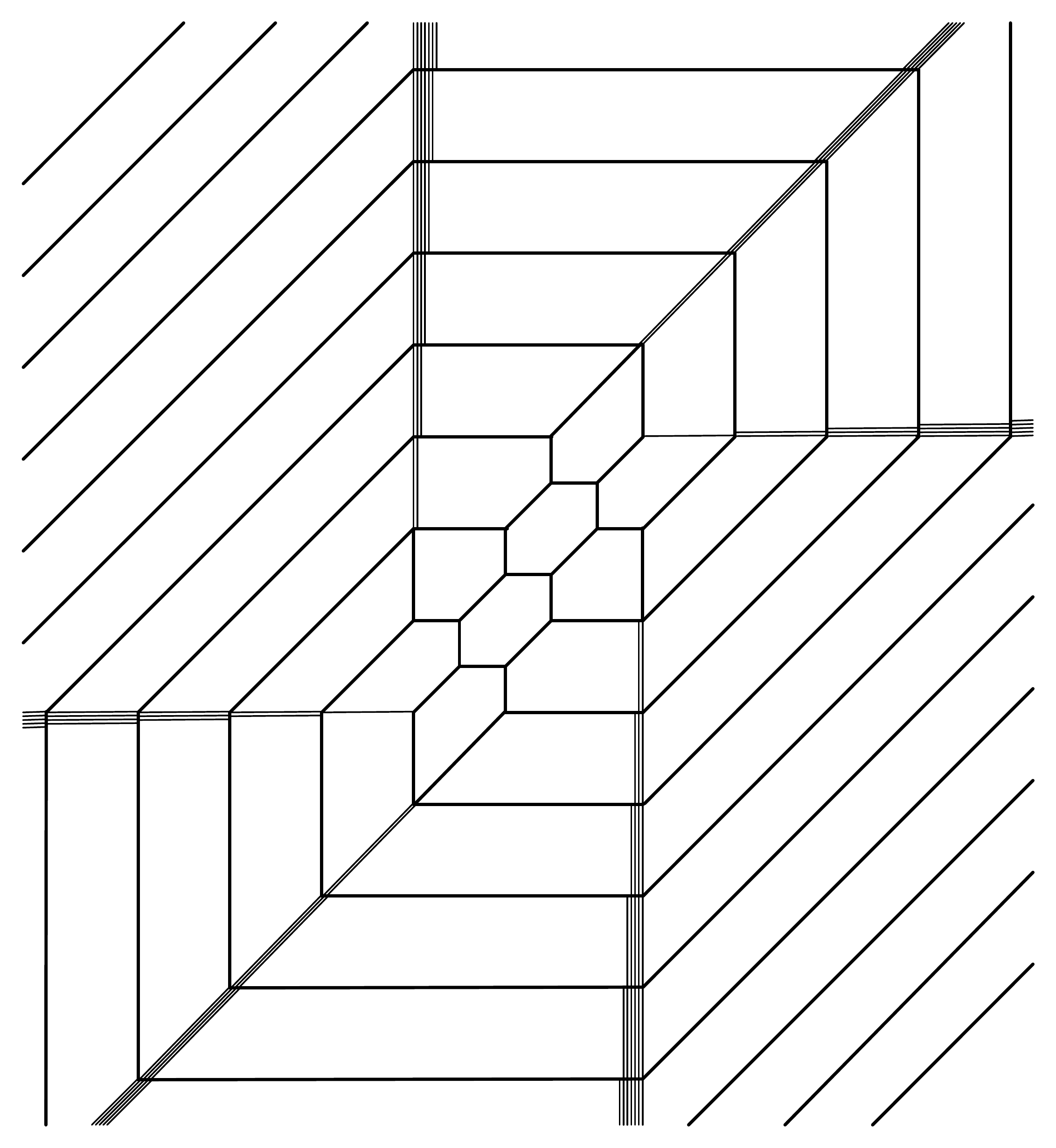}
\end{center}
\end{minipage}
\end{tabular}
\caption{Left: The 5-brane web configuration with $N_f=10$ flavors. Right: The corresponding Tao diagram which has structure of infinite spirals}
\label{fig:naive5dweb}
\end{figure}

In this and the next sections, we explicitly compute the partition functions of the 5d theories.
Here we first consider the 5d $SU(3)$ theory with $10$ flavors and compute the Nekrasov partition of the theory using the topological vertex formulation based on type IIB $(p,q)$ 5-brane web diagram.
As proposed in \cite{Hayashi:2015fsa}, 
the type IIB brane web configuration of the 6d $Sp(N-1)$ theory with $N_f=2N+4$ flavors and a tensor multiplet leads to a $(p,q)$ 5-brane web diagram for the 5d $SU(N)$ theory with $N_f=2N+4$ flavors which inherits an infinite spiral structure called {\it Tao web diagram}.

For instance, a suitable movement of 7-branes in the 
$(x_6, x_5)$-plane or the $(p, q)$-plane \cite{Kim:2015jba, Hayashi:2015fsa}
converts a naive brane configuration for such 5d theories given in  \ref{fig:naive5dweb}(a) into a Tao web diagrams shown in Figure \ref{fig:naive5dweb}(b). 
When the web diagram is reinterpreted \cite{Leung:1997tw} as a ``toric-like'' diagram \cite{Benini:2009gi, Kim:2014nqa}, 
the distances between the branes are converted into the K\"{a}hler parameters of the 
corresponding 2-cycles in the Calabi-Yau geometry.
Especially, the period of such spiral rotation in the Tao web diagram is expressed in terms of the K\"{a}hler parameters which is precisely the instanton factor $q$ \eqref{SU3instanton} obtained from the brane configuration in the previous section
\begin{align}
\prod_{n=1}^{8} \Delta^{(n)} = q^2,
\end{align}
where $\Delta^{(n)}$ are K\"{a}hler parameters associated with the distance between each arm given in Figure \ref{fig:naive5dweb}, and the precise assignment is discussed in detail in Appendix \ref{app:Kahler}. 
We note that this is a strong evidence that the Tao brane configuration captures the KK spectrum arising from a circle compactification of the 6d theory. We will show also that this is indeed true from the comparison of the two partition functions. Namely, the partition function of the 5d $SU(3)$ gauge theory with $10$ flavors, obtained based on this type IIB brane configuration, coincides with the 6d BPS partition function from the elliptic genus computation of the 6d $Sp(1)$ theory with 10 flavors and a tensor multiplet.

\subsection{The partition function}
The 5d $SU(3)$ Tao diagram is a simple generalization of that of the 5d $SU(2)$ theory with $8$ flavors \cite{Kim:2015jba}. The partition function computation of the 5d $SU(3)$ gauge theory with $10$ flavors is also similar to that of the 5d $SU(2)$ gauge theory \cite{Kim:2015jba}. For simplicity, as done in \cite{Kim:2015jba}, we restrict ourselves to the self-dual $\Omega$-background, $\epsilon_1=-\epsilon_2=\epsilon$. So we use the unrefined topological vertex formulation. 

A Tao web diagram is a complicated spiral web diagram which contains infinitely many 5-branes jumping over other 5-branes extended to infinity (external 5-branes) in the $(p,q)$-plane. 
The computational method for the web diagram including 5-branes jumping over other 5-branes are developed in \cite{Hayashi:2013qwa, Hayashi:2014wfa, Kim:2015jba, Hayashi:2015xla, Isachenkov:2014eya}.
For the unrefined case, such external 5-branes are associated with empty Young diagrams. In the computation, we cut the Tao web diagram into some sub-diagrams relevant to an expansion of the instanton factor $q$ of interest. The sub-diagram is a finite diagram where the external 5-branes are assigned to empty Young diagrams. The partition function
is then expressed in terms of an instanton expansion
\begin{align}
Z_{\text{5d SU(3)}} = \sum_k Z_k\left(g, A_1, A_2, A_3, y_1, \cdots, y_{10}\right) q^k, \label{5dSp3ptngeneric}
\end{align}
where $Z_k$ is the partition function associated with $k$ instantons, whose argument includes
the $SU(3)$ Coulomb branch moduli $A_i$ subject to $A_1A_2A_3=1$, and the mass parameters of $10$ fundamental flavors $y_i$ for $i=1, \cdots, 10$. 

\begin{figure}[t]
\centering
\includegraphics[width=14.5cm]{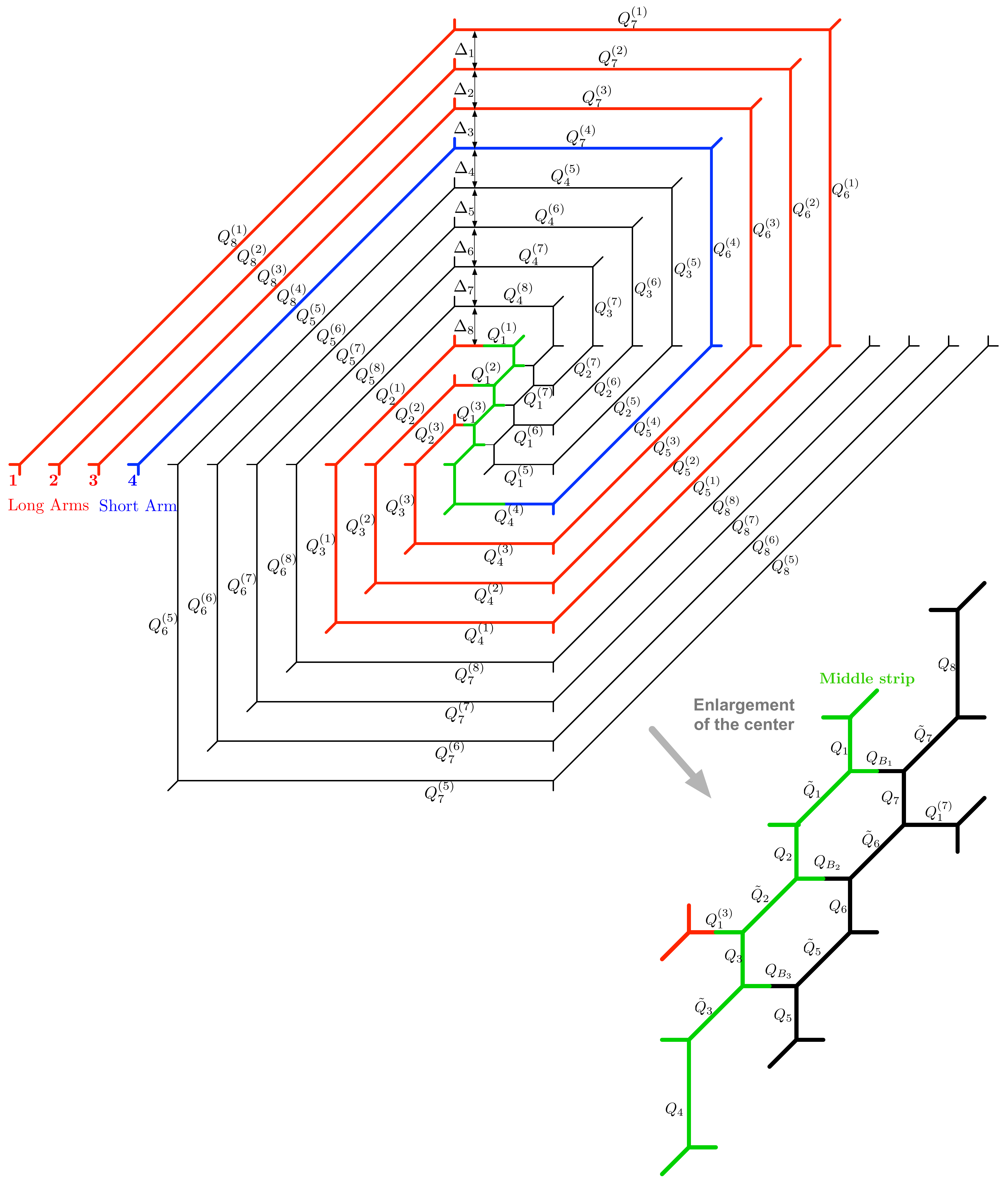}
\caption{A sub Tao diagram for 5d $SU(3)$ theory with 10 flavors relevant for the computation of the partition function up to 2 instantons. Among arms in the left, arms in red are referred to as long arms while arm in blue is referred to as short arm. The part in the middle drawn below in green is referred to the middle strip. We call the colored strips ``half 1" sub diagram, and uncolored pairs ``half 2".}
\label{fig:taodiagram}
\end{figure}

We note while the 6d elliptic genus result is computed up to 2-strings but all orders in the instanton factors, the 5d $k$-instanton partition function $Z_k$ contains all order contributions of a Coulomb branch modulus parameter which is the number of strings in the 6d elliptic genus computation. To compare these two, we double expand the partition functions in terms of the instanton factor as well as the Coulomb branch modulus parameter (or the string number). Here we compute the 5d partition function up to two instantons, $Z_2$, and also expand it in terms of $A_1$ up to the second power. Hence we need to consider a sub-diagram of Tao web diagram which has one revolution as shown in Figure \ref{fig:taodiagram}.

\begin{figure}[t]
\centering
\includegraphics[width=6cm]{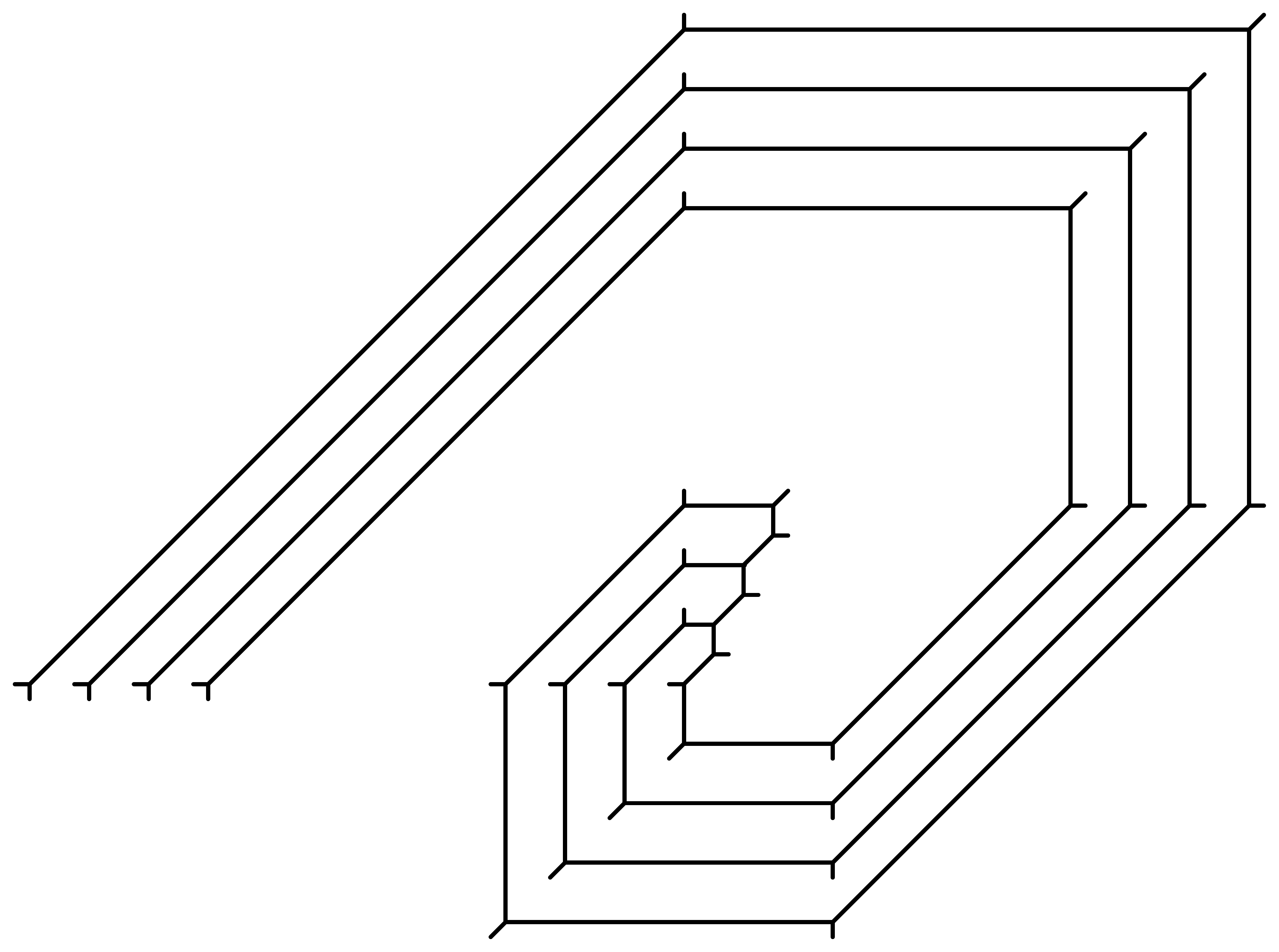}
\caption{A half Tao diagram of Figure \ref{fig:taodiagram}. This diagram is referred to  as `half1' and the other half diagram (not shown here) is referred to as `half2' for the sake of computation.}
\label{fig:halftao}
\end{figure}

It is also convenient to cut this sub Tao web diagram by half and compute only the half as the other half readily  follows from the symmetry of the web diagram. We then glue each half to make the whole diagram. See Figure \ref{fig:halftao}.
The 5d partition function then takes the following form,
\begin{align}\label{eq:5dpartitionfuntion}
Z = \sum_{Y_1, Y_2, Y_3} 
Z_{\rm glue} (Q_{B_1},Y_1) 
Z_{\rm glue} (Q_{B_2},Y_2) 
Z_{\rm glue} (Q_{B_3},Y_3) 
Z_{\rm half1} Z_{\rm half2}
\end{align}
where $Z_{\rm glue}$ is the contribution appearing when we glue some contributions
with the K\"ahler parameter $Q$ with Young diagram $Y$ associated:
\begin{align}\label{eq:Zglue}
Z_{\rm glue} (Q,Y)=
(-Q)^{|Y|} Z_Y Z_{Y^t} 
g^{ \frac{ \sum_i Y_i^2 + \sum_i Y^t_i{}^2}{2} }
\end{align}
with 
\begin{align}
Z_{\nu}(g) = \prod_{(i,j) \in \nu} (1-g^{\nu_i+\nu_j^t-i-j+1}). 
\end{align}
The amplitudes for the two sub half diagrams, $Z_{\rm half1}(y_i, ; A_j)$ and $Z_{\rm half2}(y_i, ; A_j)$, are related  by 
\begin{align}
Z_{\rm half \, 2} =  Z_{\rm half \, 1}(y_i \leftrightarrow y_{i+5}^{-1}\,; A_i \leftrightarrow A_{3-i}^{-1}).
\end{align}
Therefore, it is enough to compute only one of them, say $Z_{\rm half\,1}$.
This half amplitude consists of a strip at the middle (the middle strip) and four spiral ``arms'' (three long arms and one short arm), and it is given by, in terms of K\"ahler parameters and Young diagrams assigned in Figure \ref{fig:taodiagram}, 
\begin{align}\label{eq:Zhalf1}
Z_{\rm half \, 1} = &\sum_{\{ \mu_i \}} 
Z_{\rm middle} (\mu_1,\mu_2, \mu_3, \emptyset,Y_1,Y_2,Y_3,\mu_4,\{ Q_i\}, \{ \tilde{Q}_i \} )
\cr
&\times \Bigg[ \prod_{\ell=1}^{3}\,Z_{\rm long \, arm} (\{ Q^{(\ell)}_i \}, \mu_\ell)\Bigg]
Z_{\rm short \, arm}(\{ Q^{(4)}_i \}, \mu_4),
\end{align}
where $\mu_i$ are the Young diagram associated with the initial K\"ahler parameters $Q^{(1,2,3)}_{1}$ and $Q^{(4)}_{4}$.
The amplitude for the strip diagram is given by \cite{Iqbal:2004ne}
\begin{align}
&Z_{\rm middle} (X_1,X_2,X_3,X_4,Y_1,Y_2,Y_3,Y_4,\{Q_i, \tilde{Q}_i\})
\cr
&=\frac{
\displaystyle 
\prod_{1 \le i \le j \le 4} R_{X_i Y_j} \left( Q_j \prod_{k=i}^{j-1} Q_k \tilde{Q}_{k} \right)
\prod_{1 \le i < j \le 4} R_{Y_i X_j} \left( \tilde{Q}_i \prod_{k=i+1}^{j-1} Q_k \tilde{Q}_k \right)
}{
\displaystyle
\prod_{1 \le i \le j \le 4} R_{X_i X_j} \left( \prod_{k=i}^{j-1} Q_k \tilde{Q}_{k} \right)
\prod_{1 \le i < j \le 4} R_{Y_i Y_j} \left( \prod_{k=i}^{j-1} Q_k \tilde{Q}_{k} \right)
},
\end{align}
where 
$X_i$ and $Y_i$ are again the Young diagrams, and 
following same convention as in \cite{Bao:2013pwa}, 
\begin{align}
R_{XY}(Q) =M(Q)^{-1} \,N_{XY}(Q),
\end{align}
with 
\begin{align}
M(Q) = {\rm PE} \left[ \frac{g}{(1-g)^2} Q \right],
\end{align}
where $\rm PE$ is the Plethystic exponential defined as 
\begin{align}
{\rm PE}[f(\cdot)] = \exp \bigg[ \sum_{i=1}^{\infty} \frac1n f(\cdot^n)\bigg],
\end{align}
and 
\begin{align}
N_{\lambda \mu}(Q) = 
\prod_{(i,j) \in \lambda} (1-Q g^{\lambda_i+\mu_j^t-i-j+1} )
\prod_{(i,j) \in \mu} (1-Q g^{-\lambda_j^t-\mu_i+i+j-1} ).
\end{align}

\begin{figure}[t]
\centering
\includegraphics[width=10cm]{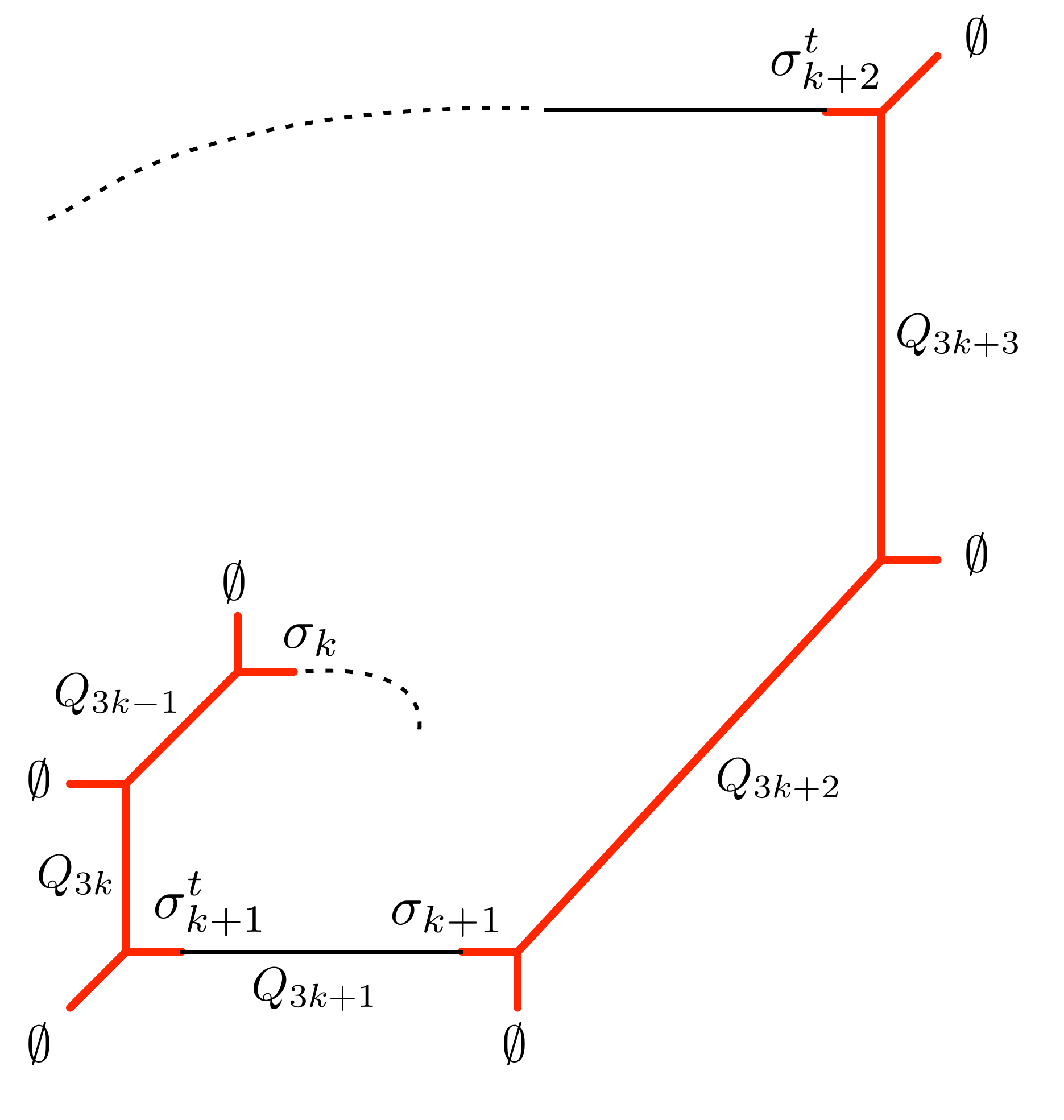}
\caption{ The strips (in red) appearing in arms of the Tao diagram.}
\label{fig:strips}
\end{figure}

The amplitude for the each ``arm'' is also computed by further decomposing into short strips depicted in Figure \ref{fig:strips}. The amplitude for each strip is given by
\begin{align}
Z_{\rm strip}(Q,Q',\sigma,\sigma'^t) = \frac{R_{\sigma \emptyset} (Q) R_{\emptyset \sigma'^t} (Q') }{R_{\sigma \sigma'^t} (Q Q')}.
\end{align}
The amplitudes for three long arms 
then have the following form:
\begin{align}\label{eq:long}
 Z_{\rm long \, arm} (\{ Q_i \}, \sigma_{1})
& = 
\sum_{ \{\sigma_i \, (i \ge 2) \}   }
\prod_{k=1}^{\infty}   
Z_{\rm glue}(Q_{3k-2}, \sigma_{k})
Z_{\rm strip}(Q_{3k-1},Q_{3k}, \sigma_k,\sigma^t_{k+1}),
\end{align} 
where $\sigma_k$ are the Young diagrams associated with $Q_{3k-2}$ and $\sigma_1$ is the initial Young diagram $\mu_\ell$ for each long arm.
The amplitude for the short arm is given by
\begin{align}\label{eq:short}
 Z_{\rm short \, arm}  (\{ Q^{(4)}_i \}, \mu_4)
& = 
\sum_{ \{\sigma_i \, (i \ge 3) \}   }
\prod_{k=2}^{\infty}  Z_{\rm glue}(Q^{(4)}_{3k-2}, \sigma_{k}) Z_{\rm strip}(Q^{(4)}_{3k-1},Q^{(4)}_{3k}, \sigma_k,\sigma^t_{k+1}),
\end{align}
where $\sigma_2=\mu_4$. 

The 5d partition function is obtained by combining the two half sub diagrams and gluing them together \eqref{eq:5dpartitionfuntion}.
We now present the result of the partition function based on the $SU(3)$ Tao diagram, order by order in the instanton expansion up to 2 instanton. For that, we use the following conventions: 
We define the ``$U(10)$" 
characters with the normalization factor $\chi_0$ as  
\begin{align}
\chi_n (y) = \chi_0\,
 \sum_{1 \le i_1 < i_2 < \cdots < i_n \le10} y_{i_1} y_{i_2} \cdots  y_{i_n} 
\qquad (n=1,2,\cdots 10),
\end{align}
where 
\begin{align}
\chi_{0} = \left( \prod_{i=1}^{10} y_i \right) {}^{-\frac{1}{2}} = \chi_{10}{}^{-1}.
\end{align}
We also introduce an ``$SU(3)$ invariant combination''
\begin{align}
P_n 
 = A_1{}^n + A_2{}^n + A_3{}^n,
\end{align}
subject to the $SU(3)$ condition $A_1 A_2 A_3 = 1$. 
In the computation, there are terms which do not explicitly depend on the Coulomb branch moduli, and by modding out such terms one reproduces the correct partition function. We call such terms the {\it extra factors} denoted by $\mathcal{E}_{n}(y)$ with $n$ being the instanton power.
Analogous claim on this extra factor is discussed in various literature including \cite{Bergman:2013ala, Hayashi:2013qwa, Bao:2013pwa, Bergman:2013aca, Hwang:2014uwa, Kim:2015jba} in various context.

\paragraph{Perturbative part:}
The perturbative contribution takes the following form
\begin{align}\label{eq:pertO}
Z_{\rm pert} = {\rm PE} \left[  \mathcal{E}_0 (y) + \mathcal{F} _0(A, y) + \mathcal{O} \big((y_4^{-1})^5, y_9{}^5\big) \right],
\end{align}
where the extra factor is 
\begin{align}
\mathcal{E}_0 (y) = \frac{g}{(1-g)^2}  
\Bigg( \sum_{1 \le i < j \le 4} y_i{}^{-1}\, y_j + \sum_{6 \le i < j \le 9} y_i\, y_j{}^{-1} \Bigg).
\end{align}
The physical part is given by 
\begin{align}\label{eq:Taopert}
\mathcal{F}_0 (y) = 
\frac{g}{(1-g)^2}  \Bigl(
&
- (y_1 + y_9 + y_{10} ) A_1{}^{-1}
\cr
&
- (y_2{}^{-1} + y_3{}^{-1}+y_4{}^{-1} + y_5{}^{-1} + y_6{}^{-1} + y_7{}^{-1} + y_8{}^{-1} ) A_1
\cr
&
- (y_1 + y_2 + y_8 + y_9 + y_{10} ) A_2 {}^{-1}
\cr
&
- (y_3{}^{-1}+y_4{}^{-1} + y_5{}^{-1} + y_6{}^{-1}+y_7{}^{-1}) A_2
\cr
&
- (y_1 + y_2 + y_3 + y_7  + y_8 + y_9 + y_{10} )  A_3{}^{-1}
\cr
&
- (y_1{}^{-1} + y_5{}^{-1} + y_6{}^{-1}) A_3
\cr
&
+ 2 (A_2 A_3{}^{-1} + A_1 A_2{}^{-1} + A_1A_3{}^{-1}  )
\Bigr).
\end{align}
If we introduce a regularization corresponding to the flop transition \cite {Iqbal:2004ne, Konishi:2006ev, Taki:2008hb} for a certain part of web diagram,
\begin{align}\label{eq:flop}
{\rm PE}\left[ \frac{g}{(1-g)^2} Q \right]
\,\, \to \,\,
{\rm PE}\left[ \frac{g}{(1-g)^2} Q^{-1} \right],
\end{align}
we obtain the perturbative contribution
\begin{align}\label{eq:SU3pert}
\mathcal{F}_0(y) = \frac{g}{(1 - g)^2}
\Big( \, (P_1 P_{-1} - 3) - P_1 \chi_0 \chi_9 \, \Big).
\end{align}
which is analogous to the known result for $N_f \le 2N$ \cite{Kim:2012gu}.
However, for later purpose, 
it is convenient to transform in such a way that
negative powers of $A_1$  
never appear 
when we eliminate $A_3$ by the traceless condition.
By further using (\ref{eq:flop}), we obtain
\begin{align}\label{eq:f0pert}
\mathcal{F}_0(y) = \frac{g}{(1 - g)^2}& 
\Bigg( 2A_1A_2^{-1}+ 2A_1A_3^{-1}+2A_2A_3^{-1}-\left( A_1+A_2\right)  \chi_0\,\chi_9 
- A_3^{-1} \,\chi_1\,\chi_{10} \,\Bigg),
\end{align}
where we note that $A_3{}^{-1} = A_1 A_2$ can be also understood as the positive power of $A_1$.

We note that the computation is reliable up to $\mathcal{O} \big((y_4^{-1})^5, y_9{}^5\big)$. It should be summed over all Young diagrams $\mu_\ell$ for the initial parts of long arms of the Tao diagram and the corresponding K\"ahler parameters $Q_1^{(\ell)}$ which explicitly depend on $y_4^{-1}$ ($\ell=1,2,3$) or $y_9$ ($\ell=5,6,7$) as written in Appendix \ref{app:Kahler}. For technical difficulty, we did not perform all order Young diagram sums, although the result should hold up to all orders. Instead, we have checked up to the second powers of $Q_1^{(\ell)}$ where $y_4^{-1}$ or $y_9$ is the expansion parameter. Likewise, the instanton contributions below are checked up to $\mathcal{O} \big((y_4^{-1})^3, y_9{}^3\big)$. See also Appendix \ref{app:subtlety} for more detail.

\paragraph{One instanton part:}
The one instanton contribution takes the form
\begin{align}\label{eq:Z1}
Z_{\rm inst} = {\rm PE} \left[  ( \mathcal{E}_1 (y) + \mathcal{F} _1(A, y) + \mathcal{O} ( (y_4{}^{-1})^{\frac{7}{2}}, y_9{}^{\frac{7}{2}}, ) q \,\right],
\end{align}
where the extra factor is given by  
\begin{align}
\mathcal{E}_1 (y)
= & 
\frac{g}{(1 - g)^2} \Bigg[
\chi_0(y) (y_1 + y_2 + y_3 + y_4) (y_6 + y_7 + y_8 + y_9) 
\cr
& +    \chi_{10} (y) (y_1{}^{-1} + y_2{}^{-1} + y_3{}^{-1} + y_4{}^{-1}) (y_6{}^{-1} + y_7{}^{-1} + y_8{}^{-1} + y_9{}^{-1}) 
\cr
& - \sqrt{\frac{y_6 y_7 y_8 y_9}{y_1 y_2 y_3 y_4}} \left( \sqrt{\frac{y_{10}}{y_{5}}} + \sqrt{\frac{y_{5}}{y_{10}}}  \right)
\Bigg].
\end{align}
The physical one instanton part is given in the following form.
\begin{align}\label{eq:f1inst}
\mathcal{F}_1(A,y)
= Z_1^{(\rm naive)} + \mathcal{F}_1^{(\rm add)}.
\end{align}
Here, we write our result in a way to compare 
with the ``naive'' $U(N)$ Nekrasov partition function
\begin{align}\label{eq:naive1}
Z_1^{(\rm naive)}
=&~ \frac{g}{(1 - g)^2}
\Bigg[
\frac{\sum_{n=0}^{10} \chi_{n} (-A_1)^{6-n}}{(A_1-A_2)^2(A_1-A_3)^2} 
 + ({\rm cyclic})
\Bigg].
\end{align}
which is obtained by formally substituting $N_f=2N+4$
to the known formula, which is actually valid for $N_f \le 2N$. 
Here ``(cyclic)'' means two more terms that are obtained by taking a cyclic permutation of $A_i$ on the first term.
Comparing our result with this,
we find that this naive formula is not valid any more for $N_f=2N+4$
but we need a following additional contribution 
\begin{align}
\mathcal{F}_1^{(\rm add)}
&=~ \frac{g}{(1 - g)^2}
\Bigg[ 
\sum_{n=0}^{2} (-1)^{n+1} 
\Big( P_1{}^{2-n} \chi_{n} + P_{-1}{}^{2-n} \chi_{10-n} \Big)
\Bigg].
\end{align}

\paragraph{Two instanton part:}
The two instanton contribution takes the form
\begin{align}\label{eq:Z2}
Z_{\rm inst} = {\rm PE} \left[  ( \mathcal{E}_2 (y) + \mathcal{F} _2 (A, y) + \mathcal{O} ((y_4{}^{-1})^3, y_9{}^3 ) {q}^2\, \right],
\end{align}
where the extra factor is given by 
\begin{align}
\mathcal{E}_2 (y)
= & \frac{g}{(1-g)^2}
\Bigg[ \frac{1}{g} + g 
+ (y_1+y_2+y_3+y_4) 
(y_1{}^{-1} + y_2{}^{-1} + y_3{}^{-1} + y_4{}^{-1} +y_5{}^{-1} + y_{10}{}^{-1})  \cr
& \quad
      + ( y_5+y_6+y_7+y_8+y_9+y_{10} ) 
      (y_6{}^{-1} + y_7{}^{-1} + y_8{}^{-1} + y_9{}^{-1} ) 
\Bigg].
\end{align}

Again, we write our two instanton result as
\begin{align}\label{eq:f2inst}
\mathcal{F}_2(A,y)
= Z_2^{\rm naive} - \frac{1}{2} (Z_1^{\rm naive}) {}^2 - \frac{1}{2} Z_1^{\rm naive} (* \to *^2 )
+ \mathcal{F}^{\rm add}  .
\end{align}
where $Z_2^{\rm naive}$ is the 2-instanton part of the naive 
$U(N)$ Nekrasov partition function
\begin{align}
Z_2^{{\rm naive}} = \frac{g^2 }{(1-g)^4 } & \sum_{m=0}^{10} \sum_{n=0}^{10} (-1)^{m+n} 
\chi_m \chi_n 
\cr
&
\Biggl[
\frac{
A_1{}^{6-m} A_2{}^{6-n}
}{
 (A_1-A_3)^2 (A_2-A_3)^2 (A_1 - A_2 g^{-1})^2 (A_1-A_2 g)^2}
\cr
& 
+ \frac{g^{8-n}  A_1{}^{12-m-n} 
}{
(1+g)^2 (A_1-A_2)^2 (A_1-A_3)^2 (A_2 - A_1 g)^2 (A_3-A_1 g)^2}
\cr
&
+ \frac{g^{-6+n} A_1{}^{12-m-n} 
}{
(1+g)^2 (A_1-A_2)^2 (A_1-A_3)^2 (A_2 - A_1 g^{-1})^2 (A_3-A_1 g^{-1})^2}
\Biggr]
\cr
& 
+ ({\rm cyclic}).
\end{align}
The second and the third term in (\ref{eq:f2inst}) is given in (\ref{eq:naive1}).
They appear since we write the Nekrasov partition function in a 
Plethystic exponential form.

Comparing our result with these, we find additional contribution,
\begin{align}
\mathcal{F}^{\rm add}
=& 
\frac{g}{(1 - g)^2}
\Biggl[
\biggl(
\frac{\sum_{m=0}^{10} \sum_{n=0}^{10} d_n (-A_1){}^{6-m} \chi_m \chi_n }{(A_1-A_2)^2 (A_1-A_3)^2}
+ ({\rm cyclic})
\biggr)
\cr
& \qquad\qquad\qquad\qquad \qquad \qquad 
+ \sum_{n=0}^{10} C_{0 n} \chi_0 \chi_n 
+ \sum_{n=1}^{10} C_{n 10} \chi_n \chi_{10} 
\Biggr] 
\end{align}
where
\begin{align}
d_0 =& - 2 A_1{}^{-1} - \frac{(1+g)^2}{g} A_1{}^2, 
\qquad 
d_1 = A_1, 
\cr
d_2 = & d_3 = \cdots =d_8 = 0,
\cr
d_9 = &A_1{}^{-1}, 
\qquad
d_{10} = - 2 A_1 -  \frac{(1+g)^2}{g} A_1{}^{-2}
\end{align}
and 
\begin{align}
&C_{00} = 
2 ( P_1{}^4 - 3 P_1{}^2 P_{-1} + P_{-1}{}^2 + 2P_1) ,
\cr
&C_{01} = - 2 ( P_1{}^3 - 2 P_1 P_{-1} + 1),
\qquad
C_{02} = 2 (P_1{}^2 - P_{-1}),
\cr
&C_{03} =  -2 P_1,
\qquad
C_{04} =  2,
\qquad
C_{05} = C_{06} = 0,
\qquad
C_{07} =-2,
\qquad
C_{08} = 2 P_{-1},
\cr
&C_{09} =  - 2 ( P_{-1}{}^2 - \frac{1}{2} P_1),
\qquad
C_{0, 10} = 2( P_1{}^3 + P_{-1}{}^3 - 3 P_1 P_{-1} - \frac{1}{2} (g+g^{-1}) ).
\end{align}
Also,
\begin{align}
C_{n 10} = C_{0,10-n} (P_{\pm 1} \to P_{\mp 1} ).
\end{align}

After removing the extra factors, we have the partition function of the 5d $SU(3)$ theory with $10$ flavors, expressed in a Plethystic exponential form,
\begin{align}\label{eq:5dSU3Nek}
Z_{\rm phys} = {\rm PE} \left[ \mathcal{F}_0 (A, y) +  \mathcal{F}_1 (A, y) q+\mathcal{F}_2 (A, y) q^2+ \mathcal{O}(q^3) \, \right],
\end{align}
where $\mathcal{F}_0$, $\mathcal{F}_1$, and $\mathcal{F}_2$ are given in \eqref{eq:f0pert}, \eqref{eq:f1inst}, and \eqref{eq:f2inst} respectively.


\subsection{Comparison with the elliptic genus of the 6d $Sp(1)$ gauge theory}\label{6dsp1eg}

We will here confirm the obtained partition function of the 5d $SU(3)$ gauge theory with $10$ flavors as well as the 5d $SU(3)$-- 6d $Sp(1)$ maps by explicitly comparing it with the elliptic genus of the 6d $Sp(1)$ gauge theory with $10$ flavors and a tensor multiplet. 
We will see that the two results precisely agree, implying that the 5d $SU(3)$ theory with $N_f=10$ hypermultiplets captures the same BPS spectrum as the 6d $Sp(1)$ gauge theory with $N_f=10$ hypermultiplets and a tensor multiplet.

The elliptic genus for the 6d $Sp(1)$ gauge theory with $10$ flavors and a tensor multiplet
is given for one and two strings as  
\footnote{We thank Joonho Kim for sharing the elliptic genus result for 2-string case which is not explicitly written in \cite{Kim:2015fxa}.} \cite{Kim:2014dza, Kim:2015fxa}
\begin{align}\label{eq:sp1ellk1}
\tilde Z_{(1)}=\frac12 \frac{\eta^2}{\theta_{1}(g)^2}\sum_{I=1}^{4}\frac{\eta^2}{\theta_I(\tilde{A}^{\pm 1})}\prod_{l=1}^{10}\frac{\theta_I(\tilde{y}_l)}{\eta},
\end{align}
and
\begin{align}\label{eq:sp2ellgenk2}
\tilde Z_{(2)} =& 
\frac{ \eta^8 }{\theta_1(g)^2 \theta_1(\tilde{A}^2 g^{\pm} )^2 \theta_1( \tilde{A}^2 )^2}
\cdot \prod_{l=1}^{10}\frac{\theta_1(\tilde{y}_l  \tilde{A}^{\pm 1})}{\eta^2}
\cr
& 
+ \sum_{I=1}^4 
\frac{\eta^8}{
2 \theta_1(g)^2 \theta_1(g^2) ^2 \theta_I(\tilde{A} g^{\pm \frac{1}{2}} )^2}
\cdot
\prod_{l=1}^{10}\frac{
\theta_I(\tilde{y}_l g^{\pm \frac{1}{2} })}{\eta^2} 
\cr
&
+ \sum_{(I,J,K) \in S}
\frac{\sigma_K \eta^8 \theta_I(1)^2
}{
4 \theta_1(g) ^4
\theta_I(g)^2 
\theta_J(\tilde{A})^2
\theta_K(\tilde{A})^2
}
\cdot\prod_{l=1}^{10}\frac{
\theta_J(\tilde{y}_l) \theta_K (\tilde{y}_l)
}{\eta^2}
\end{align}
where $S = \{ (2,2,1), (3,3,1), (4,4,1), (2,3,4),(3,4,2),(4,2,3) \}$
and $\sigma_1 = -1, \sigma_2=\sigma_3=\sigma_4=1$.
On top of that, there is an overall contribution 
to the partition function,
existing even with no self-dual string
$
\tilde{Z}_{(0)} = {\rm PE} [\tilde{\mathcal{F}}_{(0)}]
$
with
\begin{align}\label{eq:k0ell}
\tilde{\mathcal{F}}_{(0)} 
= &\frac{g}{(1-g)^2}
\left(
2 (\tilde{A}^2  + \tilde{A}{}^{-2} )
- (\tilde{A}+\tilde{A}^{-1}) \sum_{i=1}^{10} ( \tilde{y}_i + \tilde{y}_i{}^{-1}  )
\right)
\left( \frac{\tilde{q}}{1-\tilde{q}} + \frac{1}{2}\right),
\end{align}
as we explain in Appendix \ref{App:elliptic}.
Here, we omitted the term which does not depend on the Coulomb branch moduli parameter $\tilde{A}$,
which corresponds to the procedure of removing ``extra factor'' $\mathcal{E}_i$ on the 5d gauge theory side.
Writing the 6d $Sp(1)$ elliptic genus in a Pelthystic exponential form, we obtain
\begin{align}\label{eq:6dell}
\tilde{Z}_{\rm 6d \,\, Sp(1)} 
= {\rm PE} [\tilde{\mathcal{F}}_{(0)} + \tilde{\mathcal{F}}_{(1)} \phi + \tilde{\mathcal{F}}_{(2)} \phi^2 + \mathcal{O} (\phi^3)]
\end{align}
with 
\begin{align}\label{6dSp1F1F2}
\tilde{\mathcal{F}}_{(1)} = \tilde{Z}_{(1)},
\qquad
\tilde{\mathcal{F}}_{(2)} = \tilde{Z}_{(2)} - \frac{1}{2} \tilde{Z}_{(1)}{}^2 - \frac{1}{2} \tilde{Z}_{(1)}(* \to *^2). 
\end{align}

In the following, we compare the 5d $SU(3)$ Nekrasov partition function (\ref{eq:5dSU3Nek}),
which we obtained based on the Tao web diagram,
with the 6d $Sp(1)$ elliptic genus (\ref{eq:6dell}).
As discussed in section \ref{sec:maps},
the parameters between  the 6d $Sp(1)$ gauge theory and the 5d $SU(3)$ gauge theory 
are related in a non-trivial manner.
The map between the parameters used for the 6d $Sp(1)$ elliptic genus and the parameters used for the 5d $Sp(2)$ Nekrasov partition function is given 
in (\ref{6dSp1to5dSp2.map1}), (\ref{6dSp1to5dSp2.map2}) and (\ref{6dSp1to5dSp2.map3}).
Also, the relation among the parameters for the 5d $Sp(2)$ gauge theory 
and the 5d $SU(3)$ gauge theory is given in (\ref{Coulombmap}), (\ref{massmap1}), (\ref{instantonmap})
and (\ref{instantonmap}).
Combining these two, we find the map between
the 6d $Sp(1)$ gauge theory and the 5d $SU(3)$ gauge theory as
\begin{align}\label{eq:6d5dSUmap}
&\tilde{y}_i = q^{\frac{1}{2}} \left(  \prod_{j=1}^{10} y_j{}^{-\frac{1}{4}} \right) y_i, \qquad
(i=1, \cdots 5)
\cr
&\tilde{y}_i = q^{-\frac{1}{2}} \left( \prod_{j=1}^{10} y_j{}^{\frac{1}{4}} \right) y_i{}^{-1}, \qquad
(i=6, \cdots 9)
\cr
& \tilde{y}_{10} = q^{-\frac{5}{2}} \left( \prod_{j=1}^{10} y_j{}^{\frac{1}{4}} \right) y_{10}{}^{-1}, \qquad
\phi = q^2 \left( \prod_{j=1}^{10} y_j{}^{-\frac{1}{2}} \right) y_{10} A_1, \qquad
\cr
& \tilde{A} = q^{\frac{1}{2}} \left(  \prod_{j=1}^{10} y_j{}^{-\frac{1}{4}} \right) A_2, \qquad
\tilde{q}=q^2
\end{align}

Substituting this map into the 6d elliptic genus,
we find the 6d partition function,
which was originally given as an expansion in terms of the tensor branch moduli parameter $\phi$,
is now given as an expansion in terms of the 5d Coulomb branch modulus parameter $A_1$
\begin{align}
\tilde{Z}_{\text{6d Sp(1)}} &= 
{\rm PE} \Biggl[ \sum_{k=0}\tilde{\mathcal{F}}_{(k)}\left(g, \tilde{A}, \tilde{y}_i, \tilde{q} \right)\phi^k \Biggr] \cr
&=
{\rm PE} \Biggl[  \sum_{k=0} 
\left( \tilde{\mathcal{F}}_{(k)}\left(g, A_2, y_i, q \right) 
\left( q^2 y_{10} \prod_{j=1}^{10} y_j{}^{-\frac{1}{2}}  \right)^k \right)
A_1{}^k \Biggr].
\end{align}
Note that $A_1$ dependence appears only from $\phi$ 
and thus, $\tilde{\mathcal{F}}_{(k)}$ does not depend on $A_1$ after substituting the map.
Therefore, the $k$-string elliptic genus simply gives the 
coefficient of $A_1{}^{k}$ in this expansion up to the rescaling factor
$\left( q^2 y_{10} \prod_{j=1}^{10} y_j{}^{-\frac{1}{2}}\right)^k$.
Also, note that each coefficient is exact in terms of $q$.

Instead, the 5d Nekrasov partition function 
is given by an expansion in terms of the instanton factor $q$.
Each instanton contribution is exact in terms of $A_1$.
In order to compare these two with each other,
we consider the double expansion in terms of $A_1$ and $q$.
For the elliptic genus, we further expand each $\tilde{\mathcal{F}}_{(k)}$ in terms of $q$
\begin{align}
\left( q^2 y_{10} \prod_{j=1}^{10} y_j{}^{-\frac{1}{2}}  \right)^k  \tilde{ \mathcal{F} }_{(k)} 
= \sum_{\ell} \tilde{C}_{(k),\ell} \,q^\ell,
\end{align}
so that we obtain the double expansion
\begin{align}
\tilde{Z}_{\text{6d Sp(1)}} = {\rm PE} \Bigl[ \sum_{k, \ell} \tilde{C}_{(k) \ell}  A_1{}^{k}q^\ell \Bigr]
\end{align}
as a total.
For the Nekrasov partition function, we expand 
each instanton contribution in terms of $A_1$ so that we obtain the double expansion
\begin{align}
Z_{\rm 5d} 
= {\rm PE} \Bigl[ \sum_\ell \mathcal{F}_\ell q^\ell \Bigr]
= {\rm PE} \Bigl[ \sum_{k, \ell} C_{(k) \ell}  A_1{}^{k}q^\ell \Bigr].
\end{align}
Then, we compare the coefficients of these double expansion
up to $2$ instantons and $2$ strings.
Note that we eliminate 
the Coulomb branch modulus parameter $A_3$ by using the traceless condition.

By this procedure, we found the perfect agreement 
$C_{(k) \ell} = \tilde{C}_{(k) \ell}$ with $k=1,2$ and $\ell=1,2$ without any ambiguity.
We also checked that the $1$ and $2$ instanton contributions of the Nekrasov partition function
do not produce any negative power of $A_1$.
Analogously, the $1$ and $2$ string contributions of the elliptic genus do not 
produce any negative power of $q$.

The coefficients for $q^0$ and $A_1^0$ are confirmed up to 
the transition (\ref{eq:flop}).
The perturbative part of the Nekrasov partition function 
can be written in various ways up to this transition.
However, in order for this comparison to work,
we need to use (\ref{eq:f0pert}), which does not have any negative power of $A_1$, 
instead of using the original form (\ref{eq:Taopert}) obtained from the Tao diagram computation
or the well known form (\ref{eq:SU3pert}).
Analogously, instead of (\ref{eq:k0ell}), 
we need to use its ``flopped'' version 
\begin{align}\label{6dSp1flopped}
\tilde{\mathcal{F}}_{(0)} 
= &\frac{g}{(1-g)^2}
\left(
2 (\tilde{A}^2  + \tilde{A}{}^{-2} )
- (\tilde{A}+\tilde{A}^{-1}) \sum_{i=1}^{10} ( \tilde{y}_i + \tilde{y}_i{}^{-1}  )
\right)
\left( \frac{\tilde{q}}{1-\tilde{q}} \right)
\cr
& + \frac{g}{(1-g)^2}
\left( 2 \tilde{A}^2 - \tilde{A} \sum_{i=1}^{10} ( \tilde{y}_i + \tilde{y}_i{}^{-1}  )
\right),
\end{align}
which is again related to the original expression (\ref{eq:k0ell}) by the transition (\ref{eq:flop})
Under these choices, we also found the preceise agreement between the coefficients
$C_{(k) 0} = \tilde{C}_{(k) 0}$ for $k=0,1,2$
and 
$C_{(0) \ell} = \tilde{C}_{(0) \ell}$ for $\ell=0,1,2$.

\bigskip

\section{5d $Sp(2)$ gauge theory with $10$ flavors }
\label{sec:Sp2}
In this section, we devote ourselves to 5d $Sp(2)$ theory with $N_f=10$ flavors and its partition function from the point of view of the field theory rather than the brane setup. The Nekrasov instanton partition function of a $Sp(N-1)$ gauge theory was originally computed in \cite{Shadchin:2004yx,Nekrasov:2004vw} through the localization method. Especially, the 5d $Sp(N-1)$ instanton partition functions including the $O(\ell)_-$ contribution were studied in \cite{Kim:2012gu,Hwang:2014uwa} where $\ell$ denotes the instanton number. In \cite{Gaiotto:2015una}, 
 the dual relation between the 5d $Sp(N-1)$ theory with flavors and the 5d $SU(N)$ theory with the same number of fundamental hypermultiplets and a particular Chern-Simons level was analyzed from the view point of the hemisphere index. The explicit form of the partition function was computed up to $N_f\le 8$ flavors when $N=3$. 
 
However, it has not been completely understood how to calculate the partition function for the $N_f\ge 8$ case due to technical difficulties associated with higher degree poles of $O(\ell)$ holonomy integrals when the instanton number satisfies $\ell \ge2$. The one-instanton partition function, on the other hand, does not require any holonomy integral and thus the result is exact. Since the elliptic genus of the 6d $Sp(1)$ theory with $10$ flavors has the identical UV fixed point with the 5d $Sp(2)$ theory with $10$ flavors, the partition function of the 5d $Sp(2)$ theory with $10$ flavors should be the same as the result of the elliptic genus computation. We first compare the partition function up to the one-instanton order with the elliptic genus. We will then see the perfect agreement, which supports the equivalence property. Moreover, we also study the two-instanton partition function of $Sp(2)$ gauge theory and will discuss the correction form in comparison with the elliptic genus.

\subsection{The partition function up to one-instanton}
To compute the partition function for the 5d $Sp(2)$ gauge theory with $10$ fundamental flavors, we closely follow \cite{Kim:2012gu,Hwang:2014uwa} where a great detail is provided for the computation of the partition function of the 5d $Sp(N)$ gauge theory with hypermultiplets in the fundamental representation as well as hypermultiplets in the antisymmetric representation. We here simply state the result of the computation.

The computation for the partition function for 5d $Sp(2)$ gauge theory with $10$ fundamental flavors is straightforward up to one-instanton, as there is no $O(\ell)$ holonomy integral for $\ell=1$. The partition function takes the following form 
\begin{align}
Z'_{\rm 5d~Sp(2) }= 
Z'_0~ \Big( 1+  Z'_1 q + \mathcal{O}(q^2) \Big) ,\label{part.Sp2}
\end{align}
where we keep using the notation for the $\ell$-instanton partition function as $Z'_\ell$ for the 5d $Sp(2)$ gauge theory with $10$ flavors. This can also be written as a Plethystic exponential form
\begin{align}
Z_{{\rm5d~Sp(2)}}^{\prime }= {\rm PE}[ \mathcal{F}'_0 + \mathcal{F}'_1 q'+\mathcal{O}(q^{\prime 2}) ],\label{PE.Sp2}
\end{align}
where 
\begin{align}
\mathcal{F}'_0 &=
- \frac{g}{(1-g)^2}
\left( 
\sum_{i=1}^{10} (y'_i + y'_i{}^{-1}) (A'_1 + A'_2)
- 2 A'_1 (A'_2 + A'_2{}^{-1} ) 
- 2 A'_1{}^2 - 2A'_2{}^2
\right),
\label{eq:Sp2F0}\\
\mathcal{F}'_1 &= 
- \frac{g}{(1-g)^2}
\left(
\frac{
A'_1 A'_2 \prod_{i=1}^{10}(y'_i{}^{\frac{1}{2}} - y'_i{}^{-\frac{1}{2}} )
}{
(1-A'_1)^2 (1-A'_2)^2
}
+
\frac{
A'_1 A'_2 \prod_{i=1}^{10}(y'_i{}^{\frac{1}{2}} + y'_i{}^{-\frac{1}{2}} )
}{
(1+A'_1)^2 (1+A'_2)^2
}
\right),\label{eq:Sp2F1}
\end{align}
with $g=e^{-\epsilon}$ and $y'_i \; (i=1, \cdots, 10)$ are the fugacities for $SO(20)$ flavor symmetry and their chemical potentials are the $10$ flavor masses. $A'_i \; (i=1, 2)$ are the fugacities of the Cartans for the $Sp(2)$ gauge group. 
By comparing \eqref{part.Sp2} with \eqref{PE.Sp2}, it follows that 
\begin{align}
	Z_{0}^{\prime} &=~{\rm PE}\, \big[ \mathcal{F}'_0\big], \qquad Z'_1 = \mathcal{F}'_1.
\end{align}

We note that the $Sp(2)$ partition function up to one-instanton is exact in $A'_1$ and $A'_2$, while the 6d elliptic genus is exact in $q'$ as it is expressed in terms of an expansion of the string number. As done in previous section, we compare the 5d partition function with the 6d elliptic genus by doubly expanding them in the instanton fugacity $q'$ and the Coulomb modulus $A'_1$ (or the KK momentum and the string number in terms of the 6d language).  Here we expand each partition function to the quadratic order in the Coulomb branch modulus and also to the linear order in the instanton fugacity. 

The result of the 6d elliptic genus is given in section \ref{6dsp1eg}. The one-string result is \eqref{eq:sp1ellk1} and the two-string result is \eqref{eq:sp2ellgenk2}. We now implement the proposed map between the 6d $Sp(1)$ gauge theory and the 5d $Sp(2)$ gauge theory in section \ref{sec:6dSp1to5dSp2}, especially the relations\eqref{6dSp1to5dSp2.map1} and \eqref{6dSp1to5dSp2.map2}. Under the map, the 6d elliptic genus becomes
\begin{align}
\tilde{Z}_{\text{6d Sp(1)}} &=  
\text{PE}\left[\sum_{k=0}^{\infty}\tilde{\mathcal{F}}_{(k)}\left(g, \tilde{A}, \tilde{y}_1, \cdots, \tilde{y}_{10}, \tilde{q}\right)\phi^k\right]\crcr
&
=\text{PE}\left[\sum_{k=0}^{\infty}\tilde{\mathcal{F}}_{(k)}\left(g, A'_2, y'_1, \cdots, y'_9, y'_{10}q^{\prime -2}, q^{\prime 2}\right)\left(A'_1q'y^{\prime -1}_{10}\right)^k\right].
\end{align}
where $\tilde{\mathcal{F}}_{(0)}$ is given by \eqref{6dSp1flopped} 
and $ \tilde{\mathcal{F}}_{(1)}$ and $\tilde{\mathcal{F}}_{(2)}$ are given by \eqref{6dSp1F1F2}. 
This mapped elliptic genus is reorganized as a double expansion by the 5d Coulomb branch modulus $A'_1$ and also the 5d instanton fugacity $q'$,  
\begin{align}
	\tilde{Z}_{\text{6d Sp(1)}} = \text{PE}\left[\sum_{k, \ell \geq 0}\tilde{\mathcal{F}}_{(k), \ell}\left(g, A'_2, y'_i\right)A^{\prime k}_1q^{\prime \ell}\right].
\end{align}
Up to one-instanton and quadratic orders in $A'_1$, we found that the 6d elliptic genus expressed in terms of the 5d variables under the proposed map precisely coincides with the 5d one-instanton partition function. Namely we found
\begin{align}
\tilde{\mathcal{F}}_{(k), \ell}\left(g, A'_2, y'_i\right) = \mathcal{F}'_{\ell, (k)}\left(g, A'_2, y'_i\right),\label{6dSp1equal5dSp2}
\end{align}
for $0\leq \ell \leq 1$ and $0 \leq k \leq 2$, where we also expanded the 5d partition function in terms of $A'_1$ 
\begin{align}
Z_{{\rm5d~Sp(2)}}^{\prime }= {\rm PE}\left[ \sum_{k, \ell \geq 0}\mathcal{F}'_{\ell, (k)}\left(g, A'_2, y'_i\right)q^{\prime \ell}A_1^{\prime k} \right]. 
\end{align}
Note that the equality \eqref{6dSp1equal5dSp2} in the case for $k=0$ and also for $\ell=0$ is satisfied only when we use \eqref{6dSp1flopped} and \eqref{eq:Sp2F0} instead of any other choices related via (\ref{eq:flop}). 

This shows that the partition function of the 5d $Sp(2)$ gauge theory with $10$ flavors is equal to the elliptic genus of the 6d $Sp(1)$ gauge theory with $10$ flavors up to the one-instanton order and also to the two-string order. This also implies that the proposed map in section \ref{sec:6dSp1to5dSp2} is indeed the correct mapping between the two theories. 

\paragraph{Comparison between 5d $Sp(2)$ gauge theory and 5d $SU(3)$ gauge theory.}
We note here that in the previous section, the equivalence check between the partition functions of  the $Sp(2)$ gauge theory and the 5d $SU(3)$ gauge theory was done by the two successive maps: one is the map between the 6d $Sp(1)$ gauge theory and the 5d $Sp(2)$ gauge theory, which we just checked, and the other is the map between the 5d $Sp(2)$ gauge theory and the 5d $SU(3)$ gauge theory, which is based on the type IIB brane configurations. The two separate non-trivial checks here and in the previous section hence ensures that the duality relation between the 5d $Sp(2)$ gauge theory and the 5d $SU(3)$ gauge theory is also correct. 

Yet, we also consider a direct comparison between the two 5d theories. In other words, one can, in principle, obtain the partition function of the 5d $Sp(2)$ gauge theory with $10$ flavors from the partition function of the 5d $SU(3)$ gauge theory with $10$ flavors calculated by the Tao diagram, via our proposed map between the two type IIB $(p,q)$ 5-brane configurations in section \ref{sec:maps}. More precisely, the map between the two 5d theories is given in \eqref{Coulombmap}, \eqref{massmap1}, \eqref{massmap2} and also \eqref{instantonmap}. As the map involves a mixing of the instanton factor, obtaining the instanton partition function for one theory from the other is quite non-trivial. A straightforward application of the map, for example, to the one-instanton partition function gives rise to terms containing the instanton factor $q$ in the denominator, which can be reorganized as an infinite sum of the instanton factor. In order to obtain the $\ell$-instanton partition function of the 5d $Sp(2)$ gauge theory from the the instanton partition function of the 5d $SU(3)$ gauge theory, we might need all orders of the instanton partition function of the 5d $SU(3)$ gauge theory. 

The perturbative part of each, on the other hand, is quite special as can be seen from \eqref{eq:f0pert} and \eqref{eq:Sp2F0}. The application of the map gives terms up to the one-instanton factor (up to flop transitions). This implies that the perturbative part of each can be obtained from the perturbative and one-instanton part of the other theory. We indeed checked that, for instance, the perturbative part of the $Sp(2)$ partition function can be obtained from the perturbative and one-instanton part of the $SU(3)$ partition function, and vice versa.

\subsection{Two-instanton partition function}
The two-instanton partition function of the 5d $Sp(2)$ gauge theory with $10$ flavors should be treated with some caution, as  the contour integral for the case of the $N_f=10$ flavors has a triple pole at the infinity in the $O(2)$ holonomy integrals. A naive prescription of such a higher degree pole does not give any new contribution at least for the unrefined case.  Moreover, the two-instanton partition function with $10$ flavors itself has terms of negative powers of the fugacities of the Coulomb branch moduli, which implies the partition function is not well defined in a regime of very small Coulomb branch moduli. 
It is also problematic when we compare the 5d partition function with the 6d elliptic genus which is an expansion of string number fugacity $\phi$, which is related to the 5d Coulomb branch modulus $A'_1$. Therefore, in this subsection, rather than checking the agreement of the BPS partition function,
we apply the equivalence conjecture to speculate the correct 2-instanton formula.

More specifically, let us first consider the naive application of the localization for the 5d $Sp(2)$ gauge theory with $10$ flavors at the $2$-instanton level. The contribution of the naive $O(2)$ holonomy integrals for the $Sp(2)$ two-instanton partition function with $10$ flavors can be summarized as follows\footnote{As $O(k)$ has two components $O(k)_+$ and $O(k)_-$, the partition function is the sum of the two contributions, $(Z_++Z_-)/2$, where $Z_+$ has poles in the holonomy integral while $Z_- = X_5$ does not involve any contour integral. Here we will denote $X_1, X_2$ by the contributions from the poles associated with $g$ of positive and negative coefficients of $g$ respectively; $X_3, X_4$ are the contributions from the poles associated with $A'_1$ and $A'_2$ respectively.}:
\begin{align}\label{eq:naivesp2}
Z^{\prime \;{\rm naive}}_2 = \frac{1}{2} \big(X_1+X_2+X_3+X_4+X_5\big),
\end{align}
where 
\begin{align}
X_1
&= \frac{\prod_{i=1}^{10} (y'_i - g^{\frac{1}{2}}) (y_i'{}^{-1} - g^{\frac{1}{2}})
}{
(1-g)^4 (1+g)^2 (A'_1-g^{\frac{1}{2}})^2 (A'_1{}^{-1}-g^{\frac{1}{2}})^2
(A'_2-g^{\frac{1}{2}})^2  (A'_2{}^{-1} - g^{\frac{1}{2}})^2
},
\cr
X_2
&= \frac{\prod_{i=1}^{10} (y'_i + g^{\frac{1}{2}}) (y'_i{}^{-1} + g^{\frac{1}{2}})
}{
(1-g)^4 (1+g)^2 (A'_1+g^{\frac{1}{2}})^2 (A'_1{}^{-1}+g^{\frac{1}{2}})^2
(A'_2+g^{\frac{1}{2}})^2  (A'_2{}^{-1} + g^{\frac{1}{2}})^2
},
\cr
X_3
&= \frac{2 g \prod_{i=1}^{10} (y'_i -A'_1) (y'_i{}^{-1} - A'_1)
}{
(1-g)^2 A'_1{}^2 (A'_1-1)^2 (A'_1-A'_2)^2 (A'_1-A'_2{}^{-1})^2  
(A'_1{}^{2} - g)^2 (A'_1{}^{2} - g^{-1})^2
},
\cr
X_4
&= \frac{2 g \prod_{i=1}^{10} (y'_i -A_2) (y'_i{}^{-1} - A'_2)
}{
(1-g)^2 A'_2{}^2 (A'_2-1)^2 (A'_2-A'_1)^2 (A'_2-A'_1{}^{-1})^2  
(A'_2{}^{2} - g)^2 (A'_2{}^{2} - g^{-1})^2
},
\cr
X_5 
&= 
\frac{2 g^3 \prod_{i=1}^{10} (y'_i - y'_i{}^{-1})
}{
(1-g)^4 (1+g)^2 (A'_1-A'_1{}^{-1})^2 (A'_2-A'_2{}^{-1})^2
}.
\end{align}
It can be easily checked that $X_3$ has negative powers of a Coulomb branch modulus $A_1$:
\begin{align}\label{eq:negativeX3}
X_3 = 2\frac{g}{(1-g)^2} \, A_1'{}^{-2} - \frac{g}{(1-g)^2} \sum_{f=1}^{10} \left( y'_f + y'_f{}^{-1} \right)\, A_1'{}^{-1} + \mathcal{O}(A_1'{}^0).
\end{align}
Likewise, $X_4$ also contains negative powers of $A'_2$, since $X_4$ is obtained from $X_3$ by replacing  $A'_1$ with $A'_2$. As $X_3$ and $X_4$ have the negative powers of the Coulomb branch moduli, $Z_2^{\prime \;{\rm naive}}$ cannot be the correct partition function for the 5d $SU(3)$ gauge theory with $10$ flavors. 

In the previous section, we discussed that the correct partition function can be split into ``naive" and ``additional" terms. We now consider possible additional terms which may reproduce the correct Nekrasov partition function of the $2$-instanton part of the 5d $Sp(2)$ gauge theory with $10$ flavors. 
It is clear that the correct partition function should not have negative powers in Coulomb branch moduli. We thus first need to eliminate such terms of the negative powers. Given the naive result \eqref{eq:naivesp2}, we remove the terms proportional to the negative powers of the Coulomb branch moduli,  
for instance, the first two terms in \eqref{eq:negativeX3} and similar negative-power terms for $A'_2$ as well, which makes the remaining terms are of positive powers in the Coulomb branch moduli. 
We then subtract the remaining terms from the 6d elliptic genus result, which should yield the additional terms. 

To make this procedure more concrete, it is convenient to use the naive Nekrasov partition function in a Plethystic exponential form. The term associated with the two-instanton in the Plethystic exponential is given by
\begin{align}
	\mathcal{F}'_{2}{}^{\rm naive} = {Z}^{\prime\;{\rm naive}}_{2} - \frac{1}{2} {Z}_{1}{}^{\prime 2} - \frac{1}{2} Z'_{1}(* \to *^2)\, ,
\end{align}
where 
\begin{align}
\mathcal{F}'_2{}^{\rm naive}=& 
\frac{g}{(1-g)^2} 
\Bigl[
( A'_1{}^{-2}+A'_2{}^{-2})
 - \sum_{f=1}^{10} \left( y'_f + y'_f{}^{-1} \right)
(A'_1{}^{-1} +A'_2{}^{-1} )\crcr
&\qquad \qquad  + 2 (A'_1{}^{-1}A'_2{}^{-1} + A'_1A'_2{}^{-1} + A'_1{}^{-1}A'_2)+ \mathcal{O}(A_1'{}^0, A_2'{}^0)
\Bigr].
\label{eq:Sp2neg}
\end{align}
By subtracting it from the elliptic genus result \eqref{eq:6dell} taking into account the map between the 6d $Sp(1)$ gauge theory and the 5d $Sp(2)$ gauge theory, we can get the ``additional" terms. 
In other words, the correct Plethystic exponential form at the $2$-instanton order may be reproduced by
adding the ``additional" terms as in
\begin{align}\label{eq:correctSp2F2}
\mathcal{F}'_2{}^{\rm correct}= \mathcal{F}'_2{}^{\rm naive}+&
\frac{g}{(1-g)^2}
 \Bigl[
 - \chi_{\bf 4}{} ^2 + 4
 + \chi_{\bf 4}  \sum_{f=1}^{10} (y'_f+y'_f{}^{-1})
 \Bigr],
\end{align} 
where $\chi_{\bf 4}$ is the fundamental character of the $Sp(2)$ gauge group,
\begin{align}
\chi_{\bf 4} = A'_1{}^{-1} + A'_1+A'_2{}^{-1} + A'_2.
\end{align}
 
We observe that the second terms of \eqref{eq:correctSp2F2} ({\it the  additional term}) is nothing but the simplest Weyl invariant combination which can be made out of the terms of the negative powers in \eqref{eq:Sp2neg} (which of course do not introduce further negative powers). We here test our observation against the 2-instanton partition function of the 5d $Sp(1)$ theory with $8$ flavors where the correct partition function is computed by the introduction of an antisymmetric hypermultiplet \cite{Hwang:2014uwa}. Our test was done as follows: we first assume that there is no antisymmetric tensor field and compute the naive Nekrasov partition function at the 2-instanton order, which again contains negative powers of the Coulomb branch modulus $A'$ of the 5d $Sp(1)$ gauge theory. Based on the terms of the negative powers of $A'$, we make simple Weyl invariant combinations, namely $A' + A^{\prime -1}$ and $A^{\prime 2} + A^{\prime -2}$ with some functions as their coefficients, and subtract them from the naive Nekrasov partition function so that they cancel the negative powers. We then compare our proposal for the correct partition function of the 5d $Sp(1)$ gauge theory with the partition function calculated with the contribution of the antisymmetric hypermultiplet after removing extra factors. Interestingly, we found that the two results indeed agree with each other. We also found that the obtained additional terms agrees with the Higgsed version of \eqref{eq:correctSp2F2} up to extra factors which do not depend on the Coulomb branch modulus.

Note also that, for the case of the 5d $Sp(2)$ gauge theory, the result is valid up to the quadratic order in the Coulomb branch moduli, as the 6d elliptic genus is known up to that order. For the case of the 5d $Sp(1)$ gauge theory, on the other hand, the correct two-instanton partition function is known at all orders of the Coulomb branch modulus. 
We in fact found that our additional term which is a finite polynomial in $A'$ for the $Sp(1)$ case is exact. As the additional term of the 5d $Sp(1)$ gauge theory can be obtained by the Higgsing procedure on the 5d $Sp(2)$ gauge theory, it is very likely that the additional term for the $Sp(2)$ gauge theory is also exact in $A'_i$, implying that the correct two-instanton partition function given in \eqref{eq:correctSp2F2} would be exact. This also, in turn, implies that the correct two-instanton partition function for 5d $Sp(N-1)$ theory with $N_f=2N+4$ flavors is expected to have the same structure as \eqref{eq:correctSp2F2}, up to an extra factor, 
\begin{align}\label{eq:correctSpF2}
\mathcal{F}'_2{}^{\rm correct}= \mathcal{F}'_2{}^{\rm naive}\,+\,
\frac{g}{(1-g)^2}
 \Bigl[
 -\, \chi_{\bf 2N-2}{} ^2
 + \chi_{\bf 2N-2}  \sum_{f=1}^{2N+4} (y'_f+y'_f{}^{-1})
 \Bigr],
\end{align} 
where $\chi_{\bf 2N-2}$ is the character of the $(2N-2)$-dimensional representation of the $Sp(N-1)$ gauge group.

\bigskip

\section{Conclusion and discussion}
\label{sec:concl}

In this paper, we checked that the three BPS partition functions 
\begin{itemize}
\item the elliptic genus of the 6d $Sp(1)$ gauge theory with $10$ flavors and a tensor multiplet,
\item the Nekrasov partition function of the 5d $Sp(2)$ gauge theory with $10$ flavors,
\item the Nekrasov partition function of the 5d $SU(3)$ gauge theory with 10 flavors
\end{itemize}
are equivalent to one another to some orders in the expansion parameters under suitable parameter maps. This result strongly supports the original claim that the three theories have an identical UV fixed point. These theories are in an equivalence class.

Among the three BPS partition functions, only the elliptic genus had been previously computed in \cite{Kim:2015fxa}. 
For 5d $SU(3)$ and $Sp(2)$ theories, the localization method by using the conventional ADHM quantum mechanics is not simply applicable when the flavors are too many. 
For example, the perturbative part for the 5d $SU(3)$ gauge theory with $10$ flavors, and the perturbative and one-instanton parts for the 5d $Sp(2)$ gauge theory with $10$ flavors are the only contributions which one can compute from the conventional formula of the Nekrasov partition function without any difficulty.

We circumvent the difficulty by using the topological vertex formalism and applying it to the type IIB $(p, q)$ 5-brane web configuration called Tao diagram. Indeed, we have successfully computed the Nekrasov partition function of the 5d $SU(3)$ gauge theory with $10$ flavors for the first time. 
Moreover, we were also able to derive the maps among the three theories by using the 5-brane webs as well as an argument of a Higgsing.

Using the results, we compared the partition functions by 
the double expansion in terms of the instanton factor $q$ and one of the Coulomb branch moduli $A_1$, which corresponds to the string number fugacity in 6d.
Remarkably, the Nekrasov partition function of the 5d $SU(3)$ gauge theory with $10$ flavors precisely agrees with the elliptic genus of the 6d $Sp(1)$ gauge theory with $10$ flavors and a tensor multiplet
up to the $2$-instanton order, $q^2$, and the $2$ self-dual strings order, $A_1{}^2$. 
Similarly we also checked that the elliptic genus of the 6d $Sp(1)$ gauge theory with $10$ flavors and a tensor multiplet is equal to the 5d $Sp(2)$ gauge theory with $10$ flavors 
up to the $1$-instanton order, $q$, and the $2$ self-dual strings order, $A_1{}^2$. The exact agreement gives a very non-trivial evidence for the equivalence conjecture. 

As mentioned above, computing the Nekrasov partition function of the 5d $Sp(2)$ gauge theory with $10$ flavors for more than $1$-instanton has some difficulty. 
We also guessed the $2$-instanton part by assuming the equivalence.
We observed that the $2$-instanton part of the correct Nekrasov partition function obtained from the elliptic genus may be reproduced by looking into the coefficients of $A^{\prime -1}_1$ and $A_1^{\prime -2}$ of the naive Nekrasov partition function at least until the order of $A_1^{\prime 2}$. Namely, if we add the simplest Weyl invariant terms so that they cancel the terms with the negative powers of $A_1'$, the modification precisely reproduces  $2$-instanton part of the correct Nekrasov partition function of the 5d $Sp(2)$ gauge theory with $10$ flavors until the $A_1^{\prime 2}$ order. In fact, the same method turns out to be applicable to the $2$-instanton part of the 5d $Sp(1)$ gauge theory with $8$ flavors. Although the $2$-instanton part of the correct Nekrasov partition function was calculated by including the contribution of the one hypermultiplet in the antisymmetric representation in \cite{Hwang:2014uwa}, we found that it is possible to reproduce it by the $2$-instanton part of the Nekrasov partition function without the contribution of the antisymmetric hypermultiplet by adding the simplest Weyl invariant terms which cancel the contributions with negative powers of the Coulomb branch modulus of the $Sp(1)$ gauge theory. 

Although we have reproduced the $2$-instanton result of the Nekrasov partition function of the 5d $Sp(2)$ gauge theory with $10$ flavors up to the order $A_1^{\prime 2}$ by adding the simplest Weyl invariant terms which cancel negative powers of $A_1'$ from the naive Nekrasov partition function, it is not totally clear whether the same method is applied to a higher instanton order. We may need a more complicated Weyl invariant combination to cancel negative powers of a Coulomb branch modulus. Therefore, it is important to establish a more essential way to reproduce the Nekrasov partition function of the 5d $Sp(2)$ gauge theory with $10$ flavors from the localization procedure by using possibly a modified ADHM quantum mechanics in a similar manner to the method in \cite{Gaiotto:2015una}.

So far, we have focused on the case of $N=3$ for the equivalence among the 6d $Sp(N-2)$ gauge theory with $2N+4$ flavors and one tensor multiplet, the 5d $Sp(N-1)$ gauge theory with $2N+4$ flavors and the 5d $SU(N)$ gauge theory with $2N+4$ flavors. It would be interesting to further generalize our result to a different $N$. In particular, it is straightforward to generalize the topological vertex computation for the Tao diagram of the 5d $SU(N)$ gauge theory with $2N+4$ flavors for higher $N$, and the calculation will also lead to new results on the Nekrasov partition function for 5d $SU(N)$ theory with $N_f=2N+4$ flavors. 

Another example would be 6d $SU(N-1)$ theory with $N_f=N+7$ flavors and $N_a=1$ hypermultiplet in the antisymmetric representation, and one tensor multiplet. The 5d versions of the 6d $SU(N-1)$ theory are  (i) 5d $SU(N)$ gauge theory with $N_f=N+6$, $N_a=1$ hypermultiplets, and (ii) 5d quiver theory $[3n+4-N]-SU(n+1)-SU(N-n)-[2N+1-3n]$. Both have a clear IIB brane picture, which can be seen by either resolving only one $O7^-$-plane or both $O7^-$-planes. Due to the duality, these 5d theories should also have the same 6d UV fixed point and hence their Nekrasov partition functions should agree with each other. Here, with different choice of $n$, (ii) gives a collection of 5d quiver theories of different gauge group ranks and flavor symmetries, related by ``distribution duality" \cite{Hayashi:2015zka}. Especially, when $N=3$ and $n=1$, the resultant 6d/5d theories are also equivalent to what we discussed in this paper. Moreover, 5d $[4]-SU(2)-SU(2)-[4]$ quiver theory, which is the S-dual version of 5d $SU(3)$ theory with 10 flavors, may provide another approach for computing the Nekrasov instanton partition for the 5d $SU(3)$ theory.

Furthermore, there are more equivalence classes found in \cite{Hayashi:2015zka, Hayashi:2015vhy}.  An interesting equivalence class is a combination of different choice of resolution of $O7^-$-planes accompanied by $SL(2,\mathbb{Z})$ transformation. For instance, a 6d linear quiver theory $[1_A,8]-\underbrace{SU(2)-\cdots-SU(2)}_\text{$n$ nodes}-[2]$
gives rise to two quite different 5d theories: $[1_A,4] - SU(2n+1) -[1_A,4]$ and $[4] - Sp(n) - Sp(n) - [4]$. When $n=1$, it becomes the equivalent class that we considered in the paper. It would be interesting to generalize and explore other possible equivalence classes.

Although the topological vertex computation may be possible when the 5-brane web does not include O-planes, there is still a large class of Tao diagrams from which we can compute the Nekrasov partition function of various 5d theories. Some maps among the theories in an equivalence class will be also determined from a 5-brane web.


\bigskip

\acknowledgments
We thank Joonho Kim for useful discussion as well as sharing the result of the computation on the elliptic genus, and  Masato Taki for collaboration at the early stage of the work and useful discussions. The work of H.H. is supported in part by the ERC Advanced Grant SPLE under contract ERC-2012-ADG-20120216-320421, by the grant FPA2012-32828 from the MINECO, and by the grant SEV-2012-0249 of the “Centro de Excelencia Severo Ochoa” Programme. The work of K.L. is supported in part by the National Research Foundation of Korea (NRF) Grants No. 2006-0093850. We are thankful to two week long Theory Institute at CERN in February 2016, ``Recent  Developments in M-theory," supported by the CERN-Korea program.

\bigskip

\appendix 


\section{Technical details of the computation}

We here explain some technical details regarding the computation of the Nekrasov partition function for the 5d $SU(3)$ gauge theory with $10$ flavors from the Tao web diagram.  

\subsection{K\"ahler parameters of the Tao diagram}
\label{app:Kahler}
\begin{figure}[t]
\centering
\includegraphics[width=12cm]{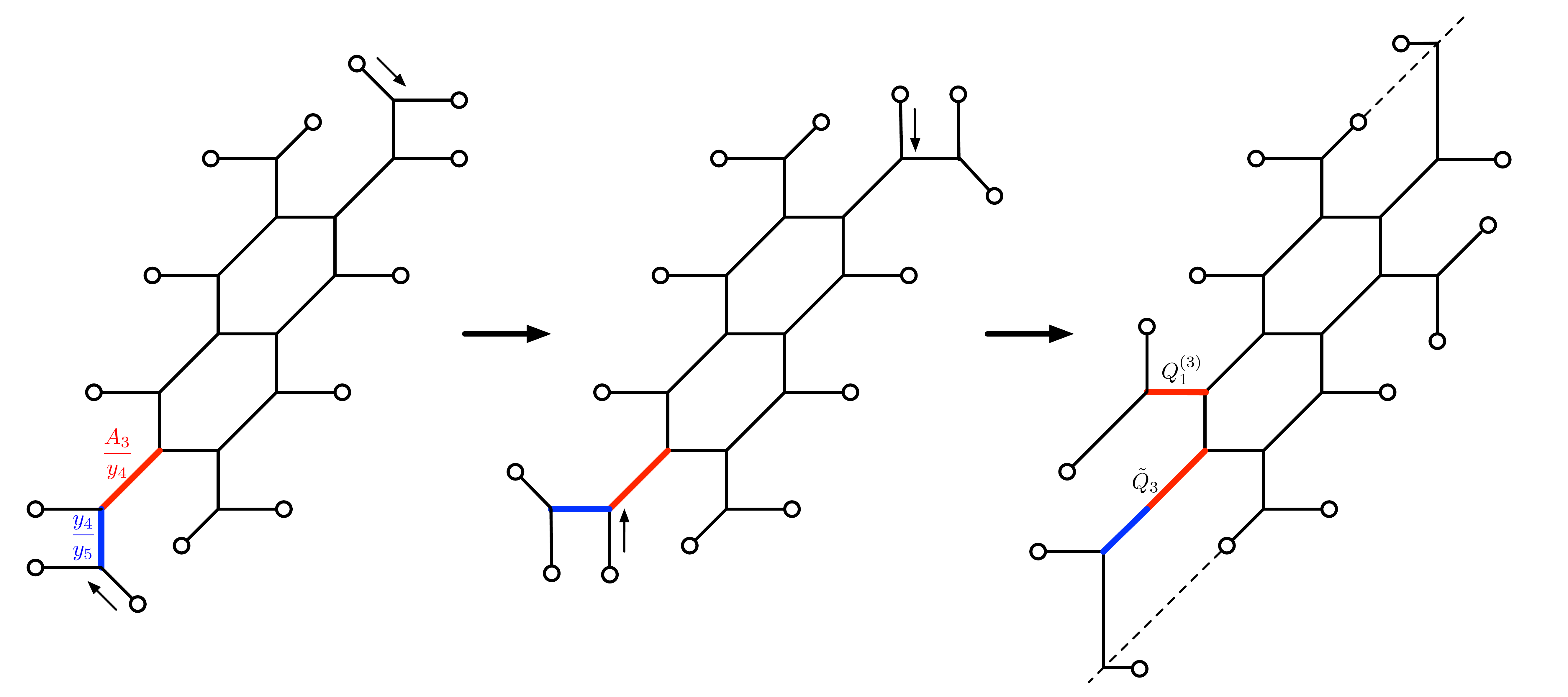}
\caption{We see that  the distance of the blue part and red part contribute to the K\"ahler parameter $\tilde{Q}_{3}$.}
\label{fig:taoderiv}
\end{figure}

Here, we summarize the expression for the K\"ahler parameters of the Tao diagram depicted in Figure \ref{fig:taodiagram} in terms of the parameters of the 5d $SU(3)$ gauge theory with $10$ flavors. 
The K\"ahler parameters are obtained by reading off the distance between the corresponding branes.
It is straightforward to read off the corresponding distances
of the middle part of the Tao diagram
by following the process to obtain 
Figure \ref{fig:naive5dweb} (Right) from 
Figure \ref{fig:naive5dweb} (Left).
\begin{align}
&Q_{1} = \frac{y_1}{A_1},\qquad
\tilde{Q}_{1} = \frac{A_1}{y_2},\qquad
Q_{2} = \frac{y_2}{A_2},\qquad
\tilde{Q}_{2} = \frac{A_2}{y_3},
\cr
&Q_{3} = \frac{y_3}{A_3},\qquad
\tilde{Q}_{3} = \frac{A_3}{y_5},\qquad
Q_{4} 
= q \sqrt{ \frac{y_4 y_5}{y_1  y_2 y_3 y_6 y_7 y_8 y_9 y_{10}} },
\cr
&Q_{5} = \frac{A_3}{y_6},\qquad
\tilde{Q}_{5} = \frac{y_7}{A_3},\qquad
Q_{6} = \frac{A_2}{y_7},\qquad
\tilde{Q}_{6} = \frac{y_8}{A_2},
\cr
&Q_{7} = \frac{A_1}{y_8},\qquad
\tilde{Q}_{7} = \frac{y_{10}}{A_1},\qquad
Q_{8} = q \sqrt{ \frac{y_6 y_7 y_8 y_1 y_2 y_3 y_4 y_{5}}{y_9 y_{10} } }.
\end{align}
For example, we see that $\tilde{Q}_{3}$
is obtained as in  Figure \ref{fig:taoderiv}.

Together with (\ref{SU3instanton}), we also obtain 
\begin{align}
&Q_{B_1} =  q \sqrt{
\frac{A_1 y_2 y_3 y_4 y_5 y_6 y_7 y_8}{y_1 y_9  y_{10}} },
\qquad
Q_{B_2} =  q \sqrt{
\frac{A_1 y_3 y_4 y_5 y_6 y_7}{A_3 y_1 y_2 y_8 y_9 y_{10} } },
\cr
&Q_{B_3} =  q \sqrt{
 \frac{y_4 y_5 y_6}{A_3 y_1y_2y_3y_7y_8y_9y_{10} } }.
\end{align}
It is convenient to define the distance $\Delta^{(i)}$ between $i$-th arm and $i+1$-th arm 
\begin{align}
&
\Delta^{(1)} = \frac{y_1}{y_2}, \quad
\Delta^{(2)} = \frac{y_2}{y_3}, \quad
\Delta^{(3)} = \frac{y_3}{y_4}, \quad
\Delta^{(4)} 
= q \sqrt{ \frac{y_4 y_6}{y_1 y_2 y_3 y_5 y_7 y_8 y_9 y_{10}} }, 
\cr
&\Delta^{(5)} = \frac{y_7}{y_6}, \quad
\Delta^{(6)} = \frac{y_8}{y_7}, \quad
\Delta^{(7)} = \frac{y_9}{y_8}, \quad
\Delta^{(8)} 
= q \sqrt{ \frac{y_2 y_3 y_4 y_5 y_6 y_7 y_8 y_{10} }{y_1 y_9} }. \quad
\end{align}
The product of the $\Delta^{(i)}$ defines the period and in fact the period precisely gives rise to the instanton fugacity for the 5d $SU(3)$ gauge theory,
\begin{align}
\prod_{n=1}^{8} \Delta^{(n)} = q^2.
\end{align}

Apart from the limited number of K\"ahler parameter with small $j$, we have 
\begin{align}
Q^{(i)}_{j} = \Delta^{(i)} Q^{(i+1)}_{j}
\qquad  j \ge 2 {\,\, \rm for \,\, } i=1,2, \qquad j \ge 3 {\,\, \rm for \,\, } i=3
\end{align}
where we identify $Q_3^{(4)} = Q_{4}$.
Therefore, it is enough to know $Q^{(4)}_j$ for  $j \ge 4$,
which is again obtained from $Q^{(5)}_{i}$
but with lower index $j$ as 
\begin{align}
Q^{(4)}_{j} = \Delta^{(4)} Q^{(5)}_{j-3}
\qquad  ( j\ge4 )
\end{align}
This $Q^{(5)}_{j-3}$ is obtained by knowing $Q^{(1)}_{j-3}$ due to the formula
\begin{align}
Q_j^{(i+4)} = Q_j^{(i)} \left( y_k \leftrightarrow y_{k+5}{}^{-1}, A_{m} \leftrightarrow A_{3-m}{}^{-1} \right)
\end{align}
which is obvious from the symmetry of the Tao diagram.
Therefore it is possible  
to compute the K\"ahler parameter $Q^{(i)}_{j}$
with large $j$ in a recursive way.

The K\"ahler parameters with small $j$,
 which is necessary to determine all the other by the recursion relation above, are given by
\begin{align}
Q^{(1)}_{1} = Q^{(2)}_{1} \tilde{Q}_{1}
\qquad  
Q^{(2)}_{1} = Q^{(3)}_{1} \tilde{Q}_{2},
\qquad
Q^{(3)}_{1} = \frac{A_3}{y_4},
\qquad
Q^{(3)}_{2} = Q_{3} \tilde{Q}_{3}.
\end{align}

The explicit expressions of all the K\"ahler parameters for the arms are as follows, 
\begin{align}
&
Q^{(1)}_1 = \frac{1}{y_2 y_3 y_4},
\quad 
Q^{(1)}_2 = \frac{y_1}{y_5},
\quad 
Q^{(1)}_3 =q \sqrt{\frac{y_1 y_5}{y_2 y_3 y_4 y_6 y_7 y_8 y_9 y_{10} }},
\quad
Q^{(1)}_4 = q \sqrt{\frac{y_1 y_6 y_7 y_8 y_9}{y_2 y_3 y_4 y_5 y_{10} }},
\cr
&
Q^{(1)}_5 =q \sqrt{\frac{y_1 y_{10}}{y_2 y_3 y_4 y_5 y_6 y_7 y_8 y_9}},
\quad
Q^{(1)}_6 = q^2 \frac{y_1}{y_{10}},
\quad 
Q^{(1)}_7 = q^2 \frac{1}{y_2 y_3 y_4},
\quad 
Q^{(1)}_8= q^2 \frac{y_1}{y_5},
\cr
&
Q^{(2)}_1 = \frac{1}{A_1 y_3 y_4},
\quad 
Q^{(2)}_2 = \frac{y_2}{y_5},
\quad 
Q^{(2)}_3 = q \sqrt{ \frac{y_2 y_5}{ y_1 y_3 y_4 y_6 y_7 y_8 y_9 y_{10} }},
\quad
Q^{(2)}_4 = q \sqrt{\frac{y_2 y_6 y_7 y_8 y_9}{y_1 y_3 y_4 y_5 y_{10} } },
\cr
&
Q^{(2)}_5 = q \sqrt{ \frac{y_2 y_{10} }{y_1 y_3 y_4 y_5 y_6 y_7 y_8 y_9} },
\quad
Q^{(2)}_6 = q^2 \frac{y_2}{y_{10}},
\quad 
Q^{(2)}_7 = q^2 \frac{1}{y_1 y_3 y_4},
\quad 
Q^{(2)}_8= q^2 \frac{y_2}{y_5},
\cr
&
Q^{(3)}_1 = \frac{A_3}{y_4},
\quad 
Q^{(3)}_2 = \frac{y_3}{y_5},
\quad 
Q^{(3)}_3 = q \sqrt{\frac{y_3 y_5}{y_1 y_2 y_4 y_6 y_7 y_8 y_9 y_{10} }},
\quad
Q^{(3)}_4 = q \sqrt{\frac{y_3 y_6 y_7 y_8 y_9}{y_1 y_2 y_4 y_5 y_{10} }},
\cr
&
Q^{(3)}_5 = q \sqrt{ \frac{y_3 y_{10} }{y_1 y_2 y_4 y_5 y_6 y_7 y_8 y_9} },
\quad
Q^{(3)}_6 = q^2 \frac{y_3}{y_{10}},
\quad 
Q^{(3)}_7 = q^2 \frac{1}{y_1 y_2 y_4},
\quad 
Q^{(3)}_8= q^2 \frac{y_3}{y_5},
\cr
&
\qquad  \qquad  \qquad  \qquad \qquad  \qquad  \qquad  
Q^{(4)}_3 = q \sqrt{ \frac{y_4 y_5}{y_1 y_2 y_3 y_6 y_7 y_8 y_9 y_{10} }},
\quad
Q^{(4)}_4 = q \sqrt{ \frac{y_4 y_6 y_7 y_8 y_9}{y_1  y_2 y_3 y_5 y_{10}} },
\cr
&
Q^{(4)}_5 =q \sqrt{ \frac{y_4 y_{10} }{y_1 y_2 y_3 y_5 y_6 y_7 y_8 y_9} },
\quad
Q^{(4)}_6 = q^2 \frac{y_4}{y_{10}},
\quad 
Q^{(4)}_7 = q^2 \frac{1}{y_1 y_2 y_3},
\quad 
Q^{(4)}_8= q^2 \frac{y_4}{y_5},
\cr
&
Q^{(5)}_1 = y_7 y_8 y_9,
\quad 
Q^{(5)}_2 = \frac{y_{10}}{y_6},
\quad 
Q^{(5)}_3 = q \sqrt{ \frac{y_1 y_2 y_3 y_4 y_5 y_7 y_8 y_9}{y_6 y_{10}} }, 
\quad
Q^{(5)}_4 = q \sqrt{ \frac{ y_5 y_7 y_8 y_9 y_{10}}{y_1 y_2 y_3 y_4 y_6} },
\cr
&
Q^{(5)}_5 = q \sqrt{ \frac{y_1 y_2 y_3 y_4 y_7 y_8 y_9 y_{10} }{y_5 y_6}},
\quad
Q^{(5)}_6 = q^2 \frac{y_5}{y_6},
\quad 
Q^{(5)}_7 = q^2 y_7 y_8 y_9,
\quad 
Q^{(5)}_8= q^2 \frac{y_{10}}{y_6},
\cr
&
Q^{(6)}_1 = A_3 y_8 y_9,
\quad 
Q^{(6)}_2 = \frac{y_{10}}{y_7},
\quad 
Q^{(6)}_3 = q \sqrt{ \frac{y_1 y_2 y_3 y_4 y_5 y_6 y_8 y_9}{y_7 y_{10}} },
\quad
Q^{(6)}_4 = q \sqrt{ \frac{y_5 y_6 y_8 y_9 y_{10} }{y_1 y_2 y_3 y_4 y_7} },
\cr
&
Q^{(6)}_5 = q \sqrt{ \frac{y_1 y_2 y_3 y_4 y_6 y_8 y_9 y_{10}}{y_5 y_7} },
\quad
Q^{(6)}_6 = q^2 \frac{y_5}{y_7},
\quad 
Q^{(6)}_7 = q^2 y_6 y_8 y_9,
\quad 
Q^{(6)}_8= q^2 \frac{y_{10}}{y_7},
\cr
&
Q^{(7)}_1 = \frac{y_9}{A_1},
\quad 
Q^{(7)}_2 = \frac{y_{10}}{y_8},
\quad 
Q^{(7)}_3 = q \sqrt{ \frac{y_1 y_2 y_3 y_4 y_5 y_6 y_7 y_9}{y_8 y_{10}} },
\quad
Q^{(7)}_4 = q \sqrt{ \frac{y_5 y_6 y_7 y_9 y_{10}}{y_1 y_2 y_3 y_4 y_8} },
\cr
&
Q^{(7)}_5 =q \sqrt{ \frac{y_1 y_2 y_3 y_4 y_6 y_7 y_9 y_{10} }{y_5 y_8} } ,
\quad
Q^{(7)}_6 = q^2 \frac{y_5}{y_8},
\quad 
Q^{(7)}_7 = q^2 y_6 y_7 y_9,
\quad 
Q^{(7)}_8= q^2 \frac{y_{10}}{y_8},
\cr
&
\qquad  \qquad  \qquad  \qquad \qquad  \qquad  \qquad  
Q^{(8)}_3 = q \sqrt{ \frac{y_1 y_2 y_3 y_4 y_5 y_6 y_7 y_8}{y_9 y_{10} } },
\quad
Q^{(8)}_4 = q \sqrt{ \frac{y_5 y_6 y_7 y_8 y_{10} }{y_1 y_2 y_3 y_4 y_9} },
\cr
&
Q^{(8)}_5 =q \sqrt{ \frac{y_1 y_2 y_3 y_4 y_6 y_7 y_8  y_{10}}{y_5 y_9} },
\quad
Q^{(8)}_6 = q^2 \frac{y_5}{y_9},
\quad 
Q^{(8)}_7 = q^2 y_6 y_7 y_8,
\quad 
Q^{(8)}_8= q^2 \frac{y_{10}}{y_9}.
\end{align}
These are all the K\"ahler parameters which are necessary to compute the $2$-instanton contribution.
We also note that the other K\"ahler parameters $Q^{(i)}_{j} $ with larger $j$ is obtained by
\begin{align}
Q^{(i)}_{j+6} = q^2 Q^{(i)}_{j} \qquad ({\rm for} \,\, i =1,2,\cdots, 8, \quad  j \ge 3).
\end{align}
The transition from $Q^{(i)}_{3n-3}$ to $Q^{(i)}_{3n-1}$ includes the instanton factor of the power $q^n$.
If we would like to compute up to $n$-instanton, we can truncate at the position of $Q^{(i)}_{3n-1}$.

\subsection{Comments on Young diagram sums}\label{app:subtlety}

Here, we discuss some computational detail of the 
topological string amplitudes of the Tao diagram.
In the computation of the 5d $SU(3)$ Nekrasov partition function,
summation over Young diagrams appears in (\ref{eq:5dpartitionfuntion}), 
(\ref{eq:Zhalf1}), (\ref{eq:long}) and (\ref{eq:short}).
These sums run for all the possible Young diagrams.
However, when we compute up to finite order of instanton,
the Young diagram sums in (\ref{eq:5dpartitionfuntion}), (\ref{eq:long}) and (\ref{eq:short}) are truncated 
at finite number of boxes 
 due to the factor of the form $Q^{|\sigma|}$ in (\ref{eq:Zglue}),
where $\sigma$ is the Young diagram and $Q$ is the K\"ahler parameter
which includes the positive power of the instanton factor.
Especially, when the K\"ahler parameter includes 
instanton factor of large enough power, only the empty Young diagram
contribute, leading to trivial contribution.
Thus, we need to consider only finite part of the Tao diagram around the center.

However, there is a difficulty in the Young diagram sum appearing in (\ref{eq:Zhalf1})
since the K\"ahler parameters $Q_i^{(1)}$ ($i=1,2,3$) does not depend on the instanton factor.
Therefore, in principle, we need to sum over all the Young diagrams
up to the ones with infinitely many boxes, which is difficult in the current computation technique.

In order to deal with this problem, we first note that 
the K\"ahler parameters $Q_i^{(1)}$ ($i=1,2,3$) all include $y_4{}^{-1}$.
Moreover, if we rescale the instanton factor as $q=p y_4{}^{-1/2}y_9{}^{1/2}$,
we find that a positive power of $y_4$ does not appear in any K\"ahler parameters.
Therefore, if we would like to compute the partition function 
up to the order of $(y_4{}^{-1})^n$,
it is enough to sum over Young diagrams associated to $Q_i^{(1)}$
 at most $n$ boxes in total.

Next, we observe the following pattern.
When we divide the partition function by the 
factor PE$[\mathcal{F}_0]$ with (\ref{eq:Taopert}),
and expand the the coefficient of $p^k$ with fixed $k$
in terms of $y_4{}^{-1}$, the expansion stops at the order of $(y_4{}^{-1})^k$.
That is, higher order terms of $y_4{}^{-1}$ all vanish.
We checked this property to the order $(y_4{}^{-1})^4$ for $p^0$,
and to the order $(y_4{}^{-1})^3$ for $q^1$ and for $q^2$.
In other words, our computation is reliable up to the terms 
$+\mathcal{O}( (y_4{}^{-1})^5)$ for $p^0$,
and $+\mathcal{O}( (y_4{}^{-1})^4$ for $q^1$ and for $q^2$.
Going back to the original parametrization of the instanton factor,
$+\mathcal{O}( (y_4{}^{-1})^5)$ for $q^0$, $+\mathcal{O}( (y_4{}^{-1})^{\frac{7}{2}})$ for $q^1$ 
and $+\mathcal{O}( (y_4{}^{-1})^3)$ for $q^2$
as written in (\ref{eq:pertO}), (\ref{eq:Z1}) and (\ref{eq:Z2}).
The situation is parallel for $y_9$.

Therefore, although we computed up to these orders,
we expect that these higher order terms actually all vanish.

\section{Zero string contribution of 6d BPS partition function}\label{App:elliptic}
The BPS partition function for 6d $\mathcal{N}=(1,0)$ $Sp(1)$ gauge theory with $10$ flavors has a 
non-trivial overall factor from ``zero-string'' contribution, 
which exists even when the self-dual string does not exist.
The zero-string contribution comes from all the multiplets, the tensor multiplet, 
the vector multiplet and the hypermultiplet.
At each level of Kaluza-Klein mode, we have the contribution 
\begin{align}
I_{\rm tensor} 
&= \frac{-t(u+u^{-1})}{(1-tu)(1-tu^{-1})},
\cr
I_{\rm vector}
& = \frac{-(1+t^2)(\tilde{A}^2+1+\tilde{A}^{-2})}{(1-tu)(1-tu^{-1})},
\cr
I_{\rm hyper} 
&= \frac{t (\tilde{A}+\tilde{A}^{-1}) \sum_{i=1}^{10} (\tilde{y}_i+\tilde{y}_i{}^{-1})}{(1-tu)(1-tu^{-1}) }, 
\end{align}
where $t=e^{-\epsilon_+}$, $u=e^{-\epsilon_-}$ with $\epsilon_{\pm} = (\epsilon_1 \pm \epsilon_2) /2$.
We should collect the contribution from all the Kaluza-Klein mode,
which is expected to generate the factor
\begin{align}
\sum_{n=1}^{\infty} \tilde{q}^n = \frac{\tilde{q}}{1-\tilde{q}}.
\end{align}
In addition to that, there are also terms which do not depend on $\tilde{q}$
\cite{Kim:2013nva}.
The zero-string contribution of 6d BPS partition function is then given by 
\begin{align}\label{eq:refinek0}
\tilde{Z}_{(0)}
=  {\rm PE} \left[
( I_{\rm tensor} + I_{\rm vector} + I_{\rm hyper} )
\left(
\frac{\tilde{q}}{1-\tilde{q}} + \frac{1}{2}
\right)
\right].
\end{align}
By dropping the contribution independent of $\tilde{A}$ (for instance, the tensor contribution, $I_{\rm tensor}$) 
and by reducing to the unrefined case by setting $t=1$ and $u=g$,
we obtain (\ref{eq:k0ell}).
We see that the term involving the factor $\frac{1}{2}$ is necessary in order to be invariant under the affine $SO(20)$ Weyl transformation up to the transition (\ref{eq:flop}).
As an example, 
when we consider the transformation
$(\tilde{y}_9,\tilde{y}_{10}) \to ( \tilde{q} \tilde{y}_{10} , \tilde{q}{}^{-1} \tilde{y}_{9})$,
we obtain 
\begin{align}
\tilde{Z}_{(0)}
\to 
\tilde{Z}_{(0)} {\rm PE}
\left[ 
\frac{(\tilde{A}+\tilde{A}^{-1}) (1+\tilde q)}{2(1-g)(1-g^{-1}) }\bigg(
-  (\tilde{y}_9{}^{-1} + \tilde{y}_{10})
+ \tilde{q}{}^{-1}  (\tilde{y}_9 + \tilde{y}_{10}{}^{-1})\bigg)
\right],
\end{align}
where the terms in PE cancel to each other when we apply (\ref{eq:flop}).


\bigskip

\providecommand{\href}[2]{#2}\begingroup\raggedright\endgroup

\end{document}